%% file: SE.tex
\documentclass[%
prd,
aps, 
reprint,
twocolumn,
superscriptaddress,
showpacs,
showkeys,
amsmath,
amssymb,
floatfix
]{revtex4-2}

\usepackage{graphicx}
\usepackage{dcolumn}
\usepackage{bm}
\usepackage{amsmath}
\usepackage{subfigure}
\usepackage{booktabs}
\usepackage{float}
\usepackage{textcomp}
\usepackage{physics}
\usepackage{xcolor}
\usepackage{siunitx}
\usepackage[colorlinks=true,citecolor=blue,filecolor=blue,linkcolor=blue,urlcolor=blue,pdftex]{hyperref}

\newcommand{\bologna}{\affiliation{Department of Physics and Astronomy, University of Bologna and INFN-Bologna, 40126 Bologna, Italy}}
\newcommand{\chicago}{\affiliation{Department of Physics \& Kavli Institute for Cosmological Physics, University of Chicago, Chicago, IL 60637, USA}}
\newcommand{\coimbra}{\affiliation{LIBPhys, Department of Physics, University of Coimbra, 3004-516 Coimbra, Portugal}}
\newcommand{\columbia}{\affiliation{Physics Department, Columbia University, New York, NY 10027, USA}}
\newcommand{\lngs}{\affiliation{INFN-Laboratori Nazionali del Gran Sasso and Gran Sasso Science Institute, 67100 L'Aquila, Italy}}
\newcommand{\mainz}{\affiliation{Institut f\"ur Physik \& Exzellenzcluster PRISMA$^{+}$, Johannes Gutenberg-Universit\"at Mainz, 55099 Mainz, Germany}}
\newcommand{\heidelberg}{\affiliation{Max-Planck-Institut f\"ur Kernphysik, 69117 Heidelberg, Germany}}
\newcommand{\munster}{\affiliation{Institut f\"ur Kernphysik, Westf\"alische Wilhelms-Universit\"at M\"unster, 48149 M\"unster, Germany}}
\newcommand{\nikhef}{\affiliation{Nikhef and the University of Amsterdam, Science Park, 1098XG Amsterdam, Netherlands}}
\newcommand{\nyuad}{\affiliation{New York University Abu Dhabi - Center for Astro, Particle and Planetary Physics, Abu Dhabi, United Arab Emirates}}
\newcommand{\purdue}{\affiliation{Department of Physics and Astronomy, Purdue University, West Lafayette, IN 47907, USA}}
\newcommand{\rice}{\affiliation{Department of Physics and Astronomy, Rice University, Houston, TX 77005, USA}}
\newcommand{\stockholm}{\affiliation{Oskar Klein Centre, Department of Physics, Stockholm University, AlbaNova, Stockholm SE-10691, Sweden}}
\newcommand{\subatech}{\affiliation{SUBATECH, IMT Atlantique, CNRS/IN2P3, Universit\'e de Nantes, Nantes 44307, France}}
\newcommand{\torino}{\affiliation{INAF-Astrophysical Observatory of Torino, Department of Physics, University  of  Torino and  INFN-Torino,  10125  Torino,  Italy}}
\newcommand{\ucsd}{\affiliation{Department of Physics, University of California San Diego, La Jolla, CA 92093, USA}}
\newcommand{\wis}{\affiliation{Department of Particle Physics and Astrophysics, Weizmann Institute of Science, Rehovot 7610001, Israel}}
\newcommand{\zurich}{\affiliation{Physik-Institut, University of Z\"urich, 8057  Z\"urich, Switzerland}}
\newcommand{\paris}{\affiliation{LPNHE, Sorbonne Universit\'{e}, Universit\'{e} de Paris, CNRS/IN2P3, 75005 Paris, France}}
\newcommand{\freiburg}{\affiliation{Physikalisches Institut, Universit\"at Freiburg, 79104 Freiburg, Germany}}
\newcommand{\napels}{\affiliation{Department of Physics ``Ettore Pancini'', University of Napoli and INFN-Napoli, 80126 Napoli, Italy}}
\newcommand{\nagoya}{\affiliation{Kobayashi-Maskawa Institute for the Origin of Particles and the Universe, and Institute for Space-Earth Environmental Research, Nagoya University, Furo-cho, Chikusa-ku, Nagoya, Aichi 464-8602, Japan}}
\newcommand{\laquila}{\affiliation{Department of Physics and Chemistry, University of L'Aquila, 67100 L'Aquila, Italy}}
\newcommand{\tokyo}{\affiliation{Kamioka Observatory, Institute for Cosmic Ray Research, and Kavli Institute for the Physics and Mathematics of the Universe (WPI), University of Tokyo, Higashi-Mozumi, Kamioka, Hida, Gifu 506-1205, Japan}}
\newcommand{\kobe}{\affiliation{Department of Physics, Kobe University, Kobe, Hyogo 657-8501, Japan}}

\newcommand{\kit}{\affiliation{Institute for Astroparticle Physics, Karlsruhe Institute of Technology, 76021 Karlsruhe, Germany}}
\newcommand{\tsinghua}{\affiliation{Department of Physics \& Center for High Energy Physics, Tsinghua University, Beijing 100084, China}}

\begin{document}

\preprint{APS/123-QED}

\title{Emission of Single and Few Electrons in XENON1T and Limits on Light Dark Matter}

\author{E.~Aprile}\columbia
\author{K.~Abe}\tokyo
\author{F.~Agostini}\bologna
\author{S.~Ahmed Maouloud}\paris
\author{M.~Alfonsi}\mainz
\author{L.~Althueser}\munster
\author{E.~Angelino}\torino
\author{J.~R.~Angevaare}\nikhef
\author{V.~C.~Antochi}\stockholm
\author{D.~Ant\'on Martin}\chicago
% ...

\collaboration{XENON Collaboration}\email[]{xenon@lngs.infn.it}

\begin{abstract}
\newpage
Delayed single- and few-electron emissions plague dual-phase time projection chambers, limiting their potential to search for light-mass dark matter. This paper examines the origins of these events in the XENON1T experiment. Characterization of the intensity of delayed electron backgrounds shows that the resulting emissions are correlated, in time and position, with high-energy events and can effectively be vetoed. In this work we extend previous S2-only analyses down to a single electron. From this analysis, after removing the correlated backgrounds, we observe rates $<$ 30\,events/(electron$\times$kg$\times$day) in the region of interest spanning 1 to 5 electrons. We derive 90\% confidence upper limits for dark matter-electron scattering, first direct limits on the electric dipole, magnetic dipole, and anapole interactions, and bosonic dark matter models, where we exclude new parameter space for dark photons and solar dark photons.

\end{abstract}

\maketitle

\input{sec_1_intro} 
\input{sec_2_data_selection}

\input{sec_3_characterization}
\input{sec_4_cuts} 
\input{sec_5_detector}
\input{sec_6_models}

\input{sec_7_conclusion}

\input{sec_8_acknowledgements}
\appendix
\input{appendixA}
\input{appendix_single_photons}

\bibliography{bibliography}

\end{document}

%% file: sec_1_intro.tex
\section{Introduction} \label{introduction}
Compelling astrophysical and cosmological evidence~\cite{Bertone:2004pz, Planck:2018vyg} for the existence of dark matter (DM) has led to numerous direct detection DM experiments~\cite{Roszkowski:2017nbc, MarrodanUndagoitia:2015veg} searching for particle candidates. Amongst those, XENON1T has set the most stringent limits on interactions of spin-independent DM masses $\geq$~3\,GeV/c$^2$~\cite{XENON:2018dbl, XENON:2019rxp, XENON:2020gfr}.

XENON1T \cite{XENON:2017aty} was a dual-phase time projection chamber (TPC) located at the INFN Laboratori Nazionali del Gran Sasso in Italy at an average depth of 3600\,m water-equivalent. The active region of the TPC contained two tonnes of liquid xenon (LXe). When a particle enters the TPC, it can scatter off a xenon nucleus or an electron, generating a nuclear recoil (NR) or electronic recoil (ER) interaction, respectively. These interactions generate scintillation photons through de-excitation of $\rm{Xe}_2$ dimers and free electrons via atomic ionization in the LXe. The scintillation signal, called the S1, is detected by two arrays of photomultiplier tubes (PMTs) located at the top and bottom of the TPC~\cite{XENON:2015ara}. An electric field between a cathode electrode (z~=~$-$97\,cm) and a gate electrode (z~=~0\,cm) is used to drift free electrons from the interaction location to the liquid-gas interface. A second, stronger field between the gate and an anode electrode extracts the electrons into the gaseous xenon (GXe) phase which then produces a secondary scintillation signal, called the S2, that is amplified proportionally to the number of extracted electrons. The three-dimensional position of the interaction location is inferred from the hit pattern of the S2 light on the top PMT array (x-y position) and from the time difference between the S1 and the S2 due to the drift time of ionization electrons (depth, z position). The S2/S1 ratio can be used to discriminate between nuclear and electronic recoils. Nuclear recoils are produced by neutron backgrounds, however they are also the primary signature for DM.   Electronic recoils are mostly from $\gamma$'s or $\beta$'s, but would also be the expected signature of leptophilic or light DM models~\cite{XENON:2017aty}. 

Interactions of particles with sub-GeV masses in the detector produce only a few prompt scintillation photons, and thus the observed S1 signals can fall below the low-energy threshold in LXe TPCs. However, due to the proportional scintillation gain in the GXe volume, the S2 signals produced by lighter particles can still be observed. Thus, to lower the energy threshold and be able to search for these lighter mass DM candidates, the requirement of having detected an S1 signal is removed. This analysis is called ``S2-only''~\cite{XENON:2019xxb}. In this work we aim to extend the previous S2-only analysis of XENON1T data~\cite{XENON:2019xxb}, which had an energy threshold of 150 photoelectrons (PE), corresponding to 5 electrons, down to below a single electron, setting our threshold at 14 PE.  

S2-only analyses increase the sensitivity for low mass candidates, but particle discrimination and z-coordinate information is lost~\cite{XENON:2019xxb}. Consequently large single- and few-electron instrumental backgrounds are observed in previous S2-only analyses~\cite{XENON10:2011prx,XENON:2016jmt,XENON:2019xxb, XENON:2019jmx, PandaX-II:2021nsg}. Lacking a full background model these analyses were limit-setting procedures only. In this work we aim to develop stringent event selections to further reduce these backgrounds and set new limits on sub-GeV DM models. In Section~\ref{data_selection} we discuss the two types of data used in XENON1T, namely continuous and triggered data, to identify single- and few-electron signals In Section~\ref{characterization}, we use both continuous and triggered data to investigate these instrumental backgrounds. As we cannot discriminate between small S2s from instrumental backgrounds and possible light DM interactions, we develop selections, described in Section~\ref{dm_cuts}, to maximize the expected signal to background ratio. In Section~\ref{sec:detector_models} we describe how we model the response of XENON1T to a given recoil energy spectrum. Despite characterization of single- and few-electron signals in this work, XENON1T lacks a full background model. Therefore in Section~\ref{sec:dm_models} we set upper limits on interaction strength on a number of DM models based on our continuous data, assuming conservatively that all surviving small S2s are signal events.

%% file: sec_2_data_selection.tex
\begin{figure*}[htb]
    \centering
    \includegraphics[width=0.98\textwidth]{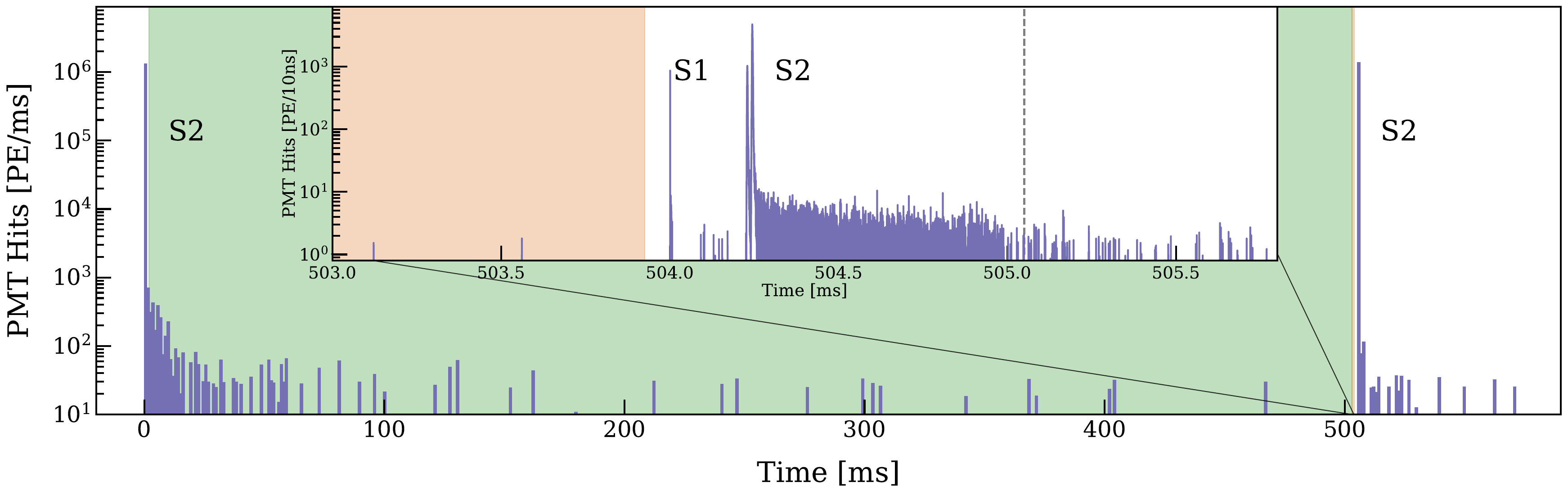}
    \caption{\textbf{Main:} An illustrative waveform in the \textit{continuous} data mode, which demonstrates the number and time distribution of few-electron S2 signals following a primary S2 signal. We define a primary S2 as an S2 having an area greater than 150\,PE. The few-electrons S2s are observed for $\mathcal{O}(100)$ times the maximum drift time of the detector. We search for few-electron signals in the continuous data between primary S2s, as indicated by the green shaded region.~\textbf{Inset:} In the \textit{triggered} data mode, only $\sim$2\,ms of data is stored. The maximum drift time in the detector is 750\,\textmu s, indicated relative to the position of the primary S2 as a vertical dashed gray line. The known background of few-electron S2s from photoionization of impurities or metal surfaces is expected to be observed within the maximum drift time. As such we search for few-electron S2s only in the orange shaded region in this data mode, before any photoionization S2s are expected. }
    \label{fig:se_events}
\end{figure*}

\section{Data} \label{data_selection}

The search for light DM interactions is complicated by the presence of a background of small S2s which can originate from $\gamma$ or $\beta$ decays of radioactive isotopes or instrumental backgrounds such as photoionization. LXe scintillation light produced by an S1 or S2 in the detector can result in photoionization electrons upon interaction with electronegative impurities or exposed metal surfaces within the detector~\cite{XENON100:2013wdu}. These electrons would be drifted to the liquid-gas interface and observed as few-electron S2s occurring within a drift time of the S1/S2. If the scintillation light from these few-electrons S2s produce additional photoionization electrons themselves, single- and few-electron signals from these electrons can be observed up to several drift times from the S1/S2. Shown in the inset of Figure~\ref{fig:se_events}, where the vertical dashed line indicates the maximum physical drift time following the S2, are the photoionization electrons following an S1/S2. These prompt photoionization S2s have been extensively studied~\cite{XENON100:2013wdu, Akerib:2020jud}, but the continued observation of few-electron S2s for $\mathcal{O}$(100x) the maximum drift time are less well understood. We therefore study isolated S2 signals smaller than $\sim$5 extracted electrons (150 photoelectrons), referred to here as few-electron signals.

In XENON1T, our standard, \textit{triggered} data acquisition software, which was designed to look for interactions with both an S1 and an S2, did not form a trigger from the smallest S2s. Thus, we can only search for potential light DM interactions that occur in close time proximity to larger S2s. Due to the low trigger rate in XENON1T, this meaningfully reduced our overall livetime as described in Section~\ref{sec:triggered_data_mode}. However, we have access to several days of \textit{continuous} data taking, described in Section~\ref{sec:triggerless_data_mode}, in which our signal detection efficiency is essentially unity for signals of one or more electrons.

\subsection{Triggered Data}\label{sec:triggered_data_mode}

The XENON1T data acquisition system (DAQ)~\cite{XENON:2019bth} was triggerless in the sense that every pulse above a $\sim$0.3 PE digitization threshold from every PMT is read out. However, XENON1T used a real-time software trigger, described in Ref.~\cite{XENON:2019bth}, to decide whether a particle interaction had occurred within the TPC. The trigger was able to distinguish S1 and S2 signals by recognizing the tight time coincidence of S1 signals relative to the broader S2 signals. In the absence of this software trigger signal, data were not stored. After classification, the trigger logic is made based on the number of pulses from individual channels that contribute to the signal. At the analysis threshold of 150\,PE used during the main science run of XENON1T~\cite{XENON:2018dbl}, referred to as SR1, the trigger accepts over 98\% of valid S2s in the center of the TPC with a background interaction rate of $\sim$5\,Hz~\cite{XENON:2019bth}. In the inset of Figure~\ref{fig:se_events} an example of the digitized data from this trigger is shown, where the trigger resulted from the S1 signal. Data is stored 1\,ms before each trigger which allows smaller S2 and S1 signals, which would not have resulted in a trigger, to be recorded. This data is then processed with the custom-developed PAX data processor~\cite{xenon_collaboration_2018_1195785}, which identifies clusters of digitized photon signals within the $\sim$2\,ms of recorded data, and classifies them as S1 or S2 signals. We will refer to this dataset as \textit{triggered} data.

In order to exclude prompt photoionization S2s, we exclude any few-electron signals which occur within 1\,ms after any S1, or any S2 greater than 150\,PE. Additionally, we exclude 75\,\textmu s prior to any trigger, as we observe that for interactions that occur within the GXe region above the anode, the S1 is occasionally mis-classified as an S2. In interactions where the mis-classified S1 is $<$~150\,PE, this signal would be considered in our few-electron population. We therefore remove this small time region prior to the trigger. We are then left with $\sim$1\,ms of data for each trigger, indicated by the orange shaded region in the inset of Figure~\ref{fig:se_events}, and a livetime of $\mathcal{O}(1\%)$.

\subsection{Continuous Data}\label{sec:triggerless_data_mode}

After the conclusion of SR1, the XENON collaboration conducted an R\&D campaign, referred to as SR2. This campaign focused on implementation and testing of a number of new systems planned for the successor experiment XENONnT~\cite{XENON:2020kmp}. One key improvement implemented during SR2 in XENON1T was a DAQ system without a software trigger. During SR2, a number of dedicated datasets were collected, using only the custom self-triggering firmware developed together with CAEN~\cite{CAEN:2021}, in which the independent trigger is implemented on each channel once the signal on that channel exceeds a configurable threshold. Data is then continuously read out and stored. Data was collected in this mode during three time periods: a one week period in February 2018, two days in April 2018 (during which data was collected under varying extraction field conditions), and one week in July 2018 (under the highest LXe purity conditions, $\sim$1\,ms electron lifetime, achieved with XENON1T). We will refer to data acquired in this mode as \textit{continuous} data. 

As with the triggered data, this data was then processed with PAX~\cite{xenon_collaboration_2018_1195785}. After interactions that deposit a large amount of energy, the high rate of observed photoionization electrons can lead to pile-up, and the resulting S2s can be in excess of our 150\,PE threshold. We therefore allow only a single S2 larger than our threshold to be identified within a set of S2s clustered by the maximum drift time in the detector. We refer to these S2s as primary S2s. In order to exclude photoionization electrons produced by the S1 of an interaction, we introduce a ``pre-trigger'' selection in continuous data, which removes 1\,ms of data (greater than the maximum drift time) before any primary S2. The resulting data window between two primary S2s is shown as a green shaded region in the main panel of Figure~\ref{fig:se_events}.

Using emulated PMT signals from a waveform simulator, we estimate the data processor to be $>$~99\,\% efficient at reconstructing S2 signals above 10\,PE. This guarantees a high acceptance for the detection of single electrons which have a gain of (28.8$\pm$0.1)\,PE measured for the science data used in this work. We apply a lower threshold of 14\,PE, where the acceptance is 99.3\%, in this analysis for both datasets.

%% file: sec_3_characterization.tex
\section{Characterization of delayed electron backgrounds} \label{characterization}

Few-electron signals in LXe TPCs can be produced by low-energy interactions due to $\gamma$ or $\beta$ decays of radioactive isotopes, instrumental backgrounds, or light DM interactions. A major instrumental background in this class of detectors is delayed electron emissions correlated in time and position with high energy interactions. This phenomenon has been observed in a variety of LXe TPCs~\cite{Akimov:2016rbs, Akerib:2020jud, XENON100:2013wdu, Bodnia:2021flk, Kopec:2021ccm}. These delayed electrons continue to be observed for $\mathcal{O}(100)$ of times the maximum drift time, as shown in the main panel of Figure~\ref{fig:se_events}.

The origin of delayed electron backgrounds in LXe TPCs has been attributed to either impurities within the TPC~\cite{Sorensen:2017kpl}, or delayed extraction of electrons at the liquid-gas interface~\cite{Sorensen:2017ymt}. In the first hypothesis for the origin of delayed emission, electrons are produced from either collisional electron detachment from negative ions, or detachment due to photoionization as a result of fluorescence of PTFE within the detector. These negative ions should form with higher probability along the drift track of the cloud of ionized electrons from an earlier interaction, and thus position correlation should be observed between the delayed few-electron signals and the prompt S2 signal of that interaction. In the second hypothesis, delayed emission is attributed to electrons trapped at the liquid-gas interface which are not emitted in the prompt S2 signal, as the extraction efficiency is not unity. These trapped electrons should also be closely matched in position with the observed prompt S2 signal. 

To understand the cause of few-electron backgrounds in XENON1T, we construct a ``delay time'' window after each primary S2, indicated by the orange shaded region for triggered data and the green shaded region for continuous data in Figure~\ref{fig:se_events}. Note that while the orange shaded region occurs before the primary S2 in triggered data, we attribute any few-electron signals to the primary S2 of the previous event. We select all few-electron signals that occur within this time window and, after correcting for the livetime of each data mode, investigate the rate as a function of interaction location within the detector (Section~\ref{sec:primary}), x-y location relative to the previous interaction (Section~\ref{sec:xy_pos}), interaction energy (Section~\ref{sec:prim_size}), time since previous interaction (Section~\ref{sec:time}), depth of the interaction that produced the primary S2 (Section~\ref{sec:drift}), as well as the overall detector conditions in terms of extraction field (Section~\ref{sec:efield}) and LXe purity (Section~\ref{purity}).

\subsection{Dependence on Interaction Location} \label{sec:primary}
\begin{figure}[ht]
    \centering
    \includegraphics[width=0.98\columnwidth]{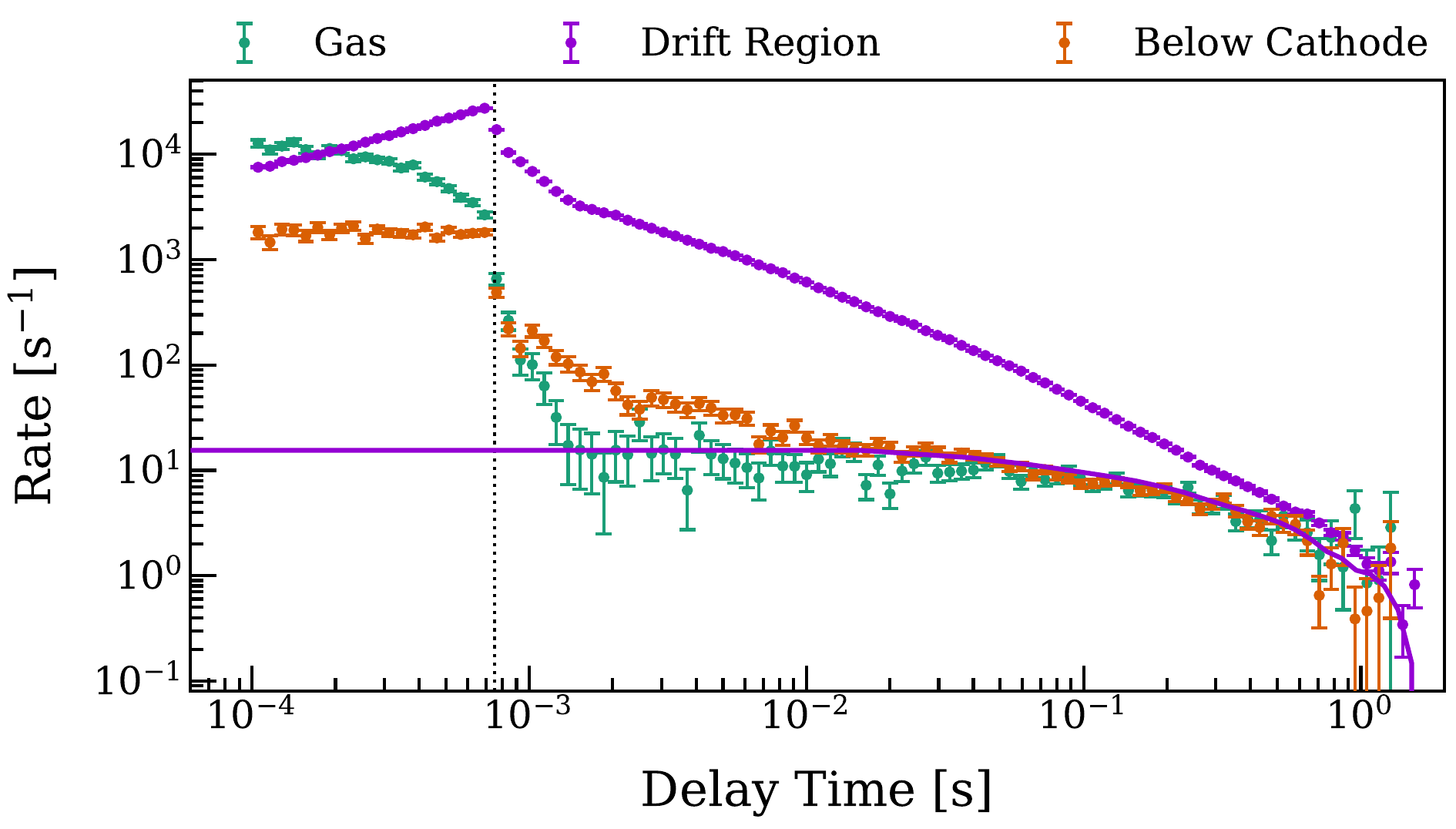}
    \caption{The rate of delayed single electron emission as a function of time following an energy deposition in the detector. The delayed emission is shown separately for interactions that occur in the GXe region (green), in the drift region between the gate and cathode electrodes (purple) and below the cathode (orange). In order to exclude delayed electrons from previous primaries, we require that no interaction has occurred for 200\,ms prior to the most recent interaction. The dotted vertical line at 750\,\textmu s indicates the maximum drift time in XENON1T for SR1 and SR2. A higher rate of delayed single electrons is observed for several times the maximum drift time. This is due to scintillation light produced as a result of the extraction of photoionization electrons into the GXe region, producing additional photoionization electrons. Shown in the purple solid line is the rate following interactions in the drift region, but shifted earlier by 200\,ms (the minimum time between interactions).}
    \label{fig:primary_dep}
\end{figure}

Interactions in the active volume of XENON1T can occur in three distinct regions: the GXe region above the liquid-gas interface (including the region above the anode), the drift region of LXe between the cathode and liquid-gas interface, and below the cathode. In the first two regions we would expect both S1 and S2 signals to be produced. However, we do not observe S2s from interactions that occur in the region below the cathode as electrons are drifted away from the GXe region by a drift field between the negatively biased cathode and a grounded screening electrode above the bottom PMT array. The two theories postulated for the origin of delayed electrons would predict that only interactions in the drift region should produce these few-electron delayed S2s~\cite{Sorensen:2017kpl,Sorensen:2017ymt}. Therefore, we select interactions from three separate regions in the detector using the continuous dataset. Interactions in the GXe region are selected by the fraction of light seen by the top PMT array. Interactions in the drift region are selected such that the time separation between the S1 and the S2 does not exceed the known maximum drift time. In the region below the cathode no S2 is observed, thus we select S1s greater than 150\,PE without an S2 within the maximum drift time. We investigate the rate of subsequent single electron emission, normalized by livetime, in the aforementioned regions, shown in Figure~\ref{fig:primary_dep}. Additionally, we require that no other interaction may have occurred in the detector within 200\,ms of the most recent interaction. We observe that for the first 750\,\textmu s (the maximum drift time) following an interaction anywhere in the detector the rate of single electron signals is high, due to photoionization of impurities within the drift region from scintillation light from either the S1 or the S2 if it is produced~\cite{XENON100:2013wdu}. At times greater than the maximum drift time in XENON1T, as seen in Figure~\ref{fig:primary_dep} for interactions in the drift region the rate drops gradually, approximately following a power law as reported in previous studies of delayed electron emission~\cite{Akerib:2020jud, Kopec:2021ccm}. By contrast, the rate of delayed electron emission following interactions in the GXe region or in the region below the cathode drops sharply and is relatively constant for $\mathcal{O}(100)$\,ms. In Figure~\ref{fig:primary_dep} the solid purple line shows the rate of delayed single electrons in the drift region, shifted earlier by 200\,ms (the minimum time between interactions). This is in agreement with the observed rate of delayed electrons from the region below the cathode and the GXe region. It is evident that interactions in the drift region are the origin of the delayed electron backgrounds in LXe TPCs and we select primary S2s from this region throughout the rest of this work. 

As the interaction rate in XENON1T science data was $\sim$\,5\,Hz, and delayed electron emission can be observed for $\mathcal{O}(100)$\,ms as shown in Figure~\ref{fig:primary_dep}, it is quite common that an interaction occurs in the delayed electron ``shadow'' of the previous interaction. After the second interaction it becomes uncertain which primary S2 a given few-electron signal should be associated with. We therefore introduce the ``S2-shadow'' parameter, $f_\mathrm{e}$, to quantify the probability that a given delayed electron is from the most recent primary S2, as opposed to a more distant antecedent primary S2. Since we observe that, as in previous studies~\cite{Akerib:2020jud, Kopec:2021ccm}, the rate of delayed electrons in XENON1T follows a power law behavior, we model the rate of delayed electrons after a primary S2 as $\propto N\cdot t^{\gamma}$, where $\gamma$ is the observed power law of the delayed electron rate following a primary S2, of which $N$ is the size in photoelectrons. We can then estimate the fraction of delayed electrons, $f_\mathrm{e}$, from the current primary S2 within a given time interval $t_i-t_f$ relative to the primary S2, as a proportion of all delayed electrons expected from all primary S2s as follows:

\begin{equation}
    f_\textrm{e} = \frac{ \int_{t_i}^{t_f} N_{m} t^{\gamma} dt} {\sum_{k \neq m} \int_{\Delta t_{km} + t_i}^{\Delta t_{km} + t_f} N_{k} t^{\gamma} dt + \int_{t_i}^{t_f} N_{m} t^{\gamma} dt} \ 
\end{equation}

\noindent where the index $m$ refers to the most recent primary S2, and the index $k$ refers to all previous primary S2s in time. $\Delta t_{km}$ is the time difference between primary S2s $k$ and $m$. For computational reasons, we restrict ourselves to only consider primary S2s within the last 2\,s when calculating $f_\mathrm{e}$, which is large relative to the $\mathcal{O}$(100)\,ms over which delayed emission is observed. The integration window is defined as being 2-200\,ms after the most recent primary S2, where the lower boundary is chosen to avoid first- and second-order photoionization electrons and the upper boundary is driven by the correspondence between data and the power law in Section~\ref{sec:time}. If the contribution of delayed electrons from all primary S2 antecedents is large compared to the contribution of delayed electrons from the current primary S2, this primary S2 and any subsequent delayed electrons up to the next primary S2 are removed from our analysis. We set our selection criteria such that interactions must fulfill $f_\mathrm{e} > 0.5$. After fitting with a power law the value of $\gamma$ is found to be $-1.04$, but small variations in the chosen value of $\gamma$ or $f_\mathrm{e}$ do not significantly alter the results presented here. Therefore we choose to conservatively round the value of $\gamma$ to $-1.0$ in the rest of this work.

\begin{figure}[t!]
    \centering
    \includegraphics[width=0.98\columnwidth]{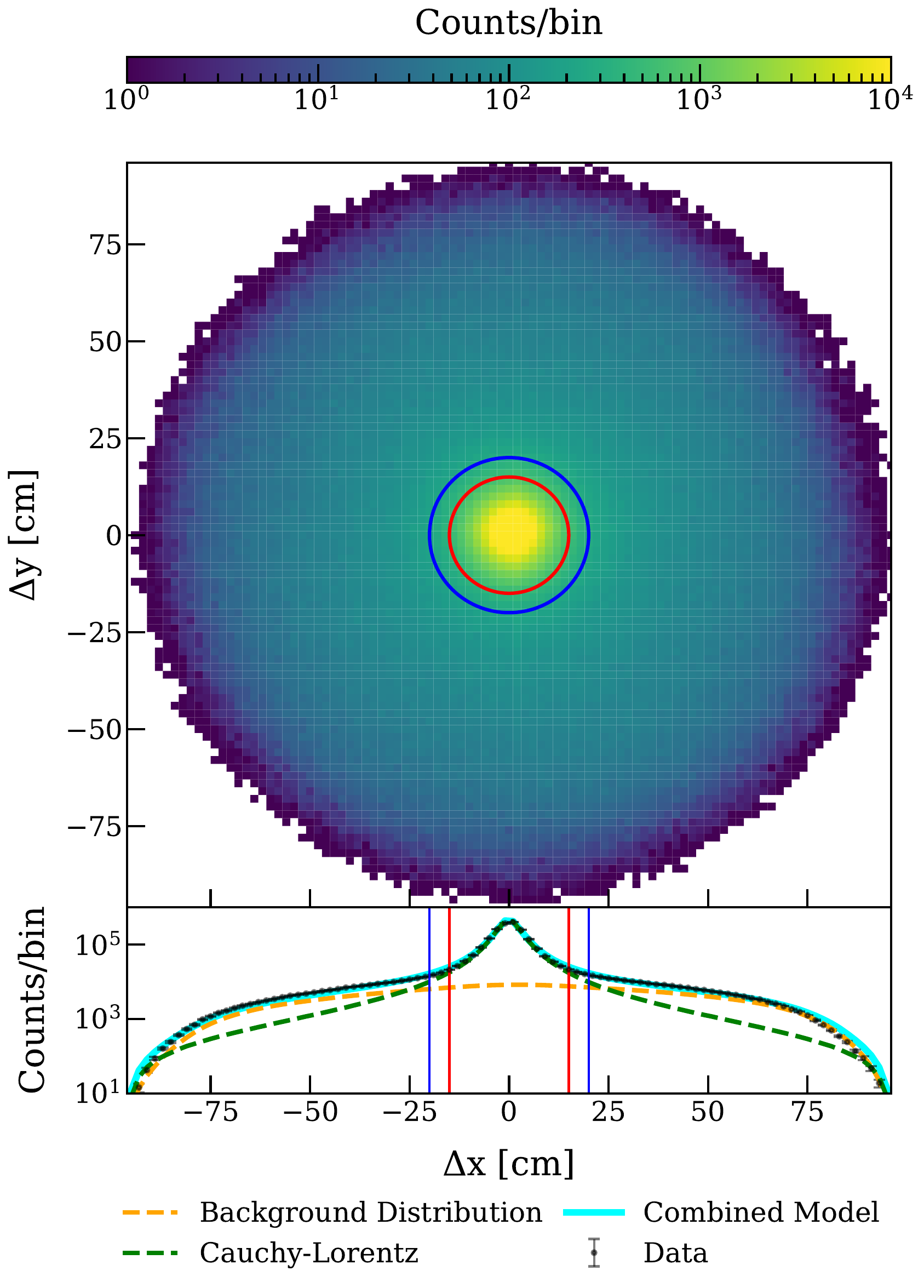}
    \caption{\textbf{Top:} Displacement of delayed single and few electrons in x-y relative to the most recent primary S2. \textbf{Bottom:} Comparison between the observed and expected displacement in the x-direction of the delayed electrons from the most recent primary S2. The expected displacement if no position correlation existed is shown by the orange dashed line, while the green dashed line represent the best fit of a Cauchy-Lorentz distribution to the correlated population. The combined model is shown in the cyan line. The red (blue) lines indicate the position correlated (uncorrelated) selection values. }
    \label{fig:dx-dy}
\end{figure}

\subsection{X-Y Position Dependence} \label{sec:xy_pos}
After applying the S2 shadow parameter and requiring the primary interaction to have occurred within the drift region, we investigate the spatial separation between the primary S2 and its subsequent delayed electrons, shown in Figure~\ref{fig:dx-dy}. Our distribution of primary S2 and delayed electron signals is not uniform across the liquid-gas interface, in both cases being strongly biased to large radii. We therefore calculate the expected displacement from random correlation of the primary S2s and delayed electrons in the triggered dataset, shown in the lower panel of Figure~\ref{fig:dx-dy} in the orange dashed line. At small values of the displacement ($\Delta \mathrm{x}$ or $\Delta \mathrm{y}$), we observe a clear excess in data indicating position correlation between the primary S2 and the majority of the delayed electron emission. This excess cannot be described by a Gaussian distribution due to the S2-size dependent position reconstruction of both the primary S2 and the delayed electron broadening the tails of the distribution. Thus we choose to describe the excess by a Cauchy-Lorentz distribution shown in the green dashed line in Figure~\ref{fig:dx-dy}. The width of the Cauchy-Lorentz distribution, defined as the half width at half maximum, is consistent with the known position reconstruction uncertainty for single electrons in XENON1T~\cite{XENON:2019ykp}. We observe that the width of the Cauchy-Lorentz model, at times greater than 10\,ms, is (2.9$\pm$0.1)\,cm and does not depend on the time since the primary S2.

As the region of increased emissions at small displacements from the primary S2 is small relative to the total area of the detector in the x-y plane, the total rate of electron emission can be comparable across the rest of the detector. Therefore, using the combined model shown in the cyan line in Figure~\ref{fig:dx-dy}, we define our region of ``position correlated'' electrons to have occurred $<$~15\,cm from the primary S2, where the boundary is chosen such that $>$~99\% of delayed electrons within the chosen radii are attributable to the Cauchy-Lorentz distribution as indicated in the bottom panel of Figure~\ref{fig:dx-dy} by the vertical red lines. 

\begin{table}[h]
\centering
\begin{tabular}{l r r r r r}
\toprule
&  \multicolumn{5}{c}{$\Delta \mathrm{r}$ [cm]} \\
\cline{2-6} & 10 & 20 & 30 & 40 & 50 \\
\hline
1e$^-$ & 47\% & 33\% & 25\% & 21\% & 21\% \\
2e$^-$ & 43\% & 28\% & 22\% & 20\% & 19\%  \\
3-5e$^-$ & 17\% & 10\% & 7\% & 6\% & 6\%  \\
\bottomrule
\end{tabular}
\caption{Percentage of delayed electron signals that occur at a radial distance $\Delta \mathrm{r}$ or greater from the preceding primary S2 that are attributed to the correlated electron population. The percentage is calculated by comparing the integrated value of the Cauchy-Lorentz model (shown by the green dashed line in Figure~\ref{fig:dx-dy}) to that of the Combined model (shown by the cyan line in Figure~\ref{fig:dx-dy}) at radial distances greater than $\Delta \mathrm{r}$.}
\label{table:purity}
\end{table}

We divide our population of delayed electron signals into 1 electron (14-42\,PE), 2 electron (42-70\,PE), and 3-5 electron (70-150\,PE) selections and perform fits on each population. The selection boundaries are chosen such that leakage from one population into another are minimized, based on fitting a model consisting of 5 Gaussian distributions to spectrum of few-electron signals below 150 PE. Shown in Table~\ref{table:purity} is the percentage of delayed electrons that are attributed to the correlated electron population for distances greater than $\Delta \mathrm{r}$ as a function of the radial distance $\Delta \mathrm{r}$ from the preceding primary S2. For each of the 1, 2 and 3-5 delayed electron populations we compare the integrated value of the Cauchy-Lorentz model to that of the Combined model at radial distances greater than $\Delta \mathrm{r}$. We observe that even at large values of $\Delta \mathrm{r}$, for the 1 and 2 electron population the fraction of correlated electrons represent a non-negligible proportion of the total number of delayed electrons. We therefore choose to define ``position uncorrelated'' electrons as delayed electron that occurred $>$~20\,cm from the primary S2. This value, indicated by the blue lines in Figure~\ref{fig:dx-dy}, is chosen such that our sensitivity to DM models is maximized. The wide spread of the Cauchy-Lorentz distribution results in our ``uncorrelated'' electrons retaining a fraction of correlated electrons as shown in Table~\ref{table:purity}. We apply these selections in subsequent sections to investigate these populations separately, correcting for the relevant detector surface area in the plane of the liquid-gas interface in each region. 

\subsection{Dependence on Interaction Size} \label{sec:prim_size}
\begin{figure}[t!]
    \centering
     \includegraphics[width=0.98\columnwidth]{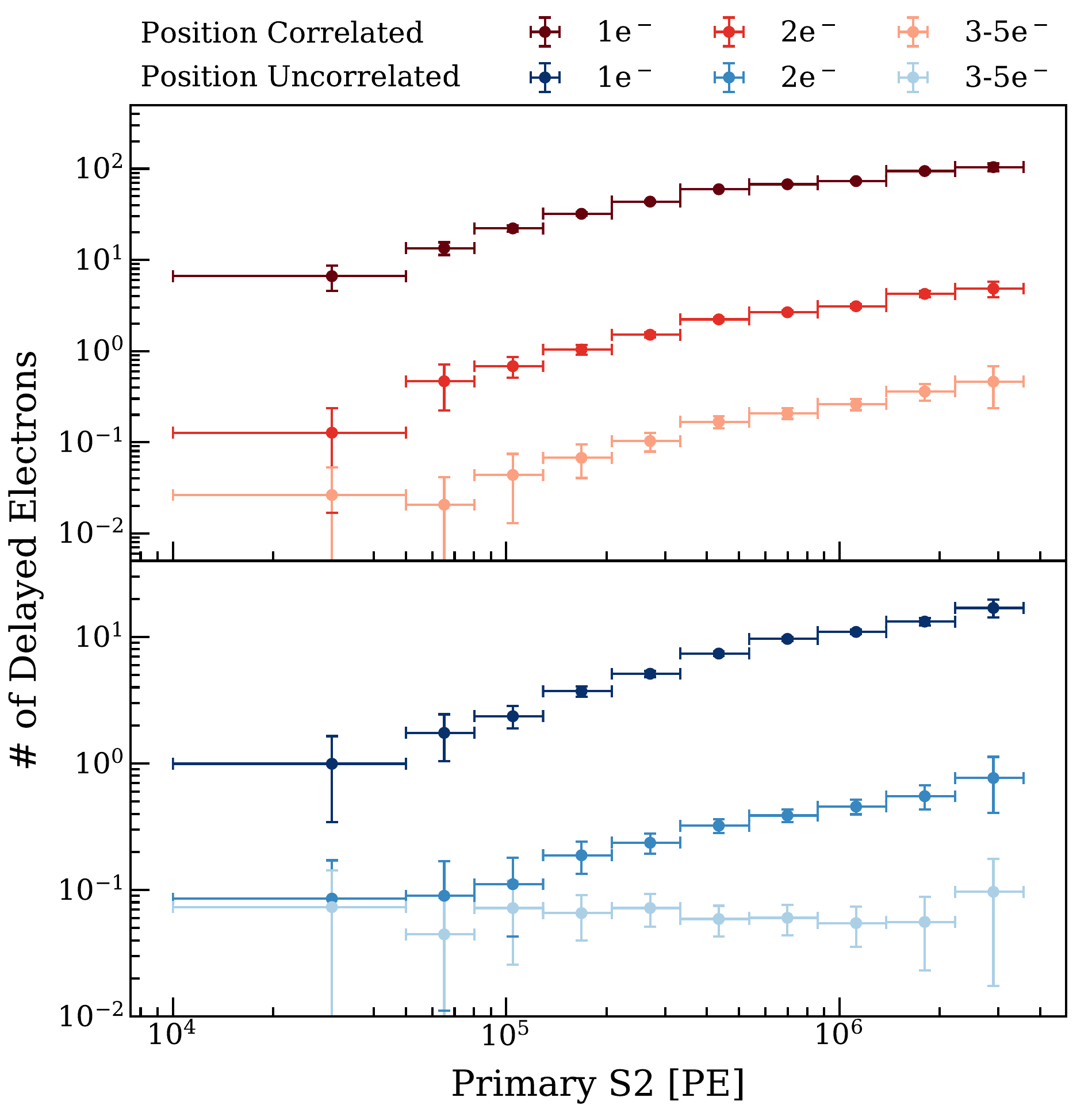}
    \caption{The number of delayed electrons observed from 2-200\,ms after the primary S2 as a function of the size of the primary S2 in photoelectrons. We observe a general trend of more delayed electrons following larger primary S2s, both for position correlated ($\Delta \mathrm{r}$~$<$~15\,cm, red in top figure) and position uncorrelated ($\Delta \mathrm{r}$~$>$~20\,cm, blue in bottom figure) delayed electrons. The rates are shown for 1 electron (dark), 2 electron (medium), and 3-5 electron (light) bins. %Additionally shown is the number of delayed electrons observed from 2-200\,ms after the primary S2 as a function of primary S2 size if a more aggressive position-uncorrelated selection is used ($\Delta \mathrm{r}$~$>$~50\,cm, grey in bottom figure).
    }
    \label{fig:prim_size}
\end{figure}

The spectrum of primary S2 interactions, as defined in this work, spans many orders of magnitude, from five electrons, at the chosen threshold of the analysis, to several hundred thousand electrons for the most energetic interactions. The origin of delayed electrons has been hypothesized to originate either from delayed extraction of electrons at the liquid-gas interface or from impurities with the LXe. Therefore, we investigate the number of delayed electrons observed as a function of the size of the primary S2 (Figure~\ref{fig:prim_size}). We use the continuous dataset and select all delayed electrons that occur within 2-200\,ms after a primary S2.  We find that as the size of the primary S2 increases so does the number of delayed electrons. The relationship between the primary S2 size and the number of delayed electrons is however not observed to be directly proportional. For position uncorrelated electrons we still see an increasing trend with larger primary S2, however we observe fewer delayed electrons in comparison to the correlated populations. For the 1 electron uncorrelated population the increase in the number of delayed electrons from the smallest to the largest primary S2s is similar to that observed for correlated single electrons, while for the 2 electron uncorrelated population this trend is slightly smaller than for position correlated electrons. 
%The increase in the uncorrelated population is due to residual contamination by correlated few-electrons, and can be ameliorated by requiring a larger separation between the few-electron signal and the primary S2 (as shown in Table~\ref{table:purity} and Figure~\ref{fig:prim_size} bottom). For example, increasing the $\Delta \mathrm{r}$ requirement to 50\,cm from 20\,cm reduces the number of uncorrelated electrons for the smallest primary S2s by a factor 1.6 and for the largest primary S2s by more than a factor of 2. The overall rate of correlated 3-5 electrons is much smaller, and thus contamination in the uncorrelated 3-5 electron population is much less severe. 

\begin{figure}[t!]
    \centering
    \includegraphics[width=0.98\columnwidth]{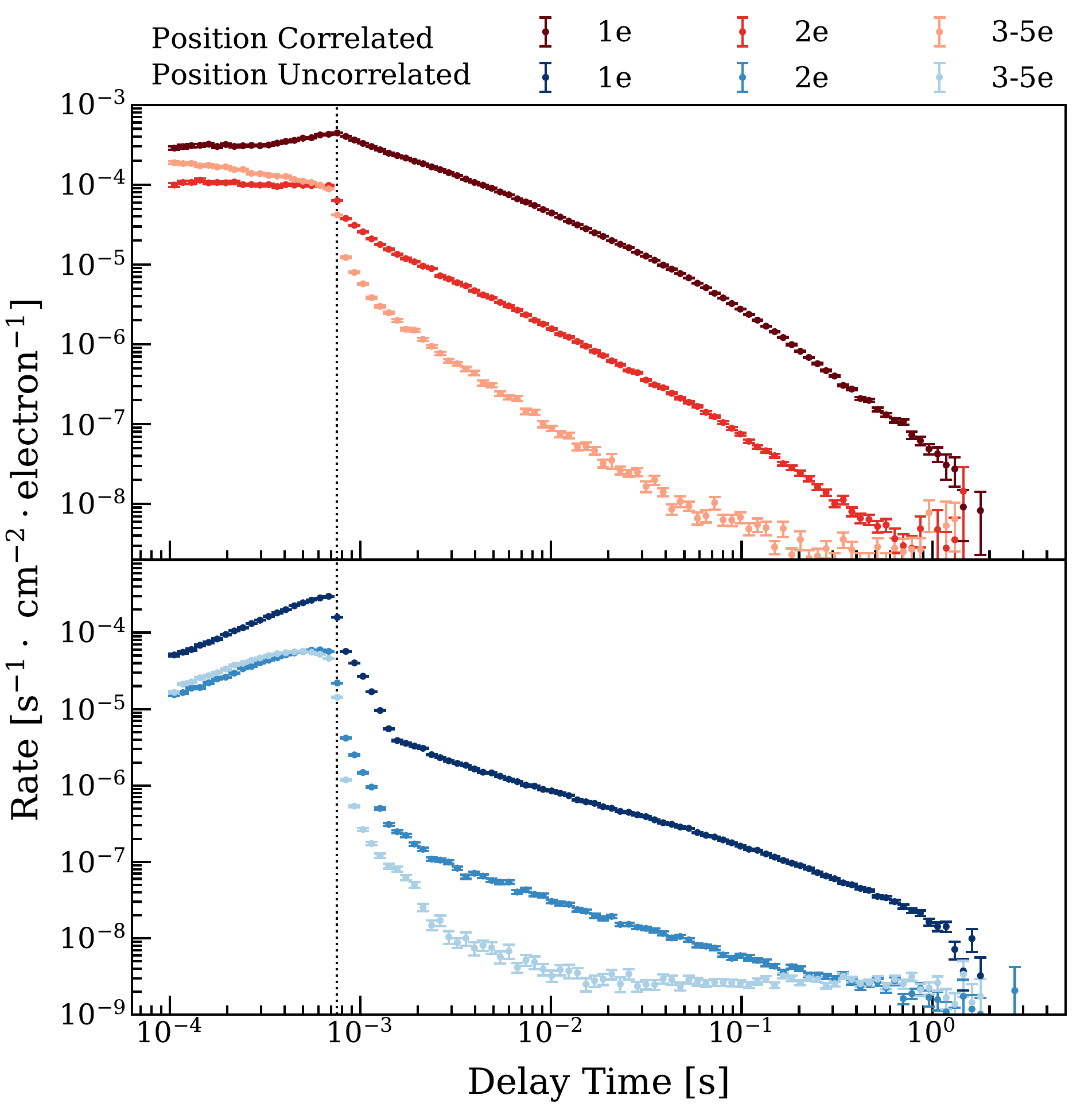}
    \caption{\textbf{Top (Bottom):} Delayed electron emission rate of position correlated (uncorrelated) electrons, normalized by detector area and number of electrons in the primary S2. Position correlated (uncorrelated) electrons are required to be displaced from the primary S2 by less (more) than 15\,cm (20\,cm). The rates are shown for the 1 electron (dark), 2 electron (medium), and 3-5 electron (light) populations. The dotted vertical line indicates the maximum drift time in XENON1T for SR1 and SR2.}
    \label{fig:corr_vs_uncorr}
\end{figure}

\subsection{Temporal Dependence} \label{sec:time}
As shown in Figure~\ref{fig:primary_dep}, the rate of delayed electron emission approximately follows a power law behavior. The observed emission rate of each of the 1 electron, 2 electron and 3-5 electron populations are shown in Figure~\ref{fig:corr_vs_uncorr} for the correlated (uncorrelated) electrons in the top (bottom) panel using the continuous dataset. Additionally, we normalize for the number of electrons in the primary S2, as we observe a dependence of the delayed electron emission rate on the size of the primary S2, discussed in Section~\ref{sec:prim_size}. We fit a power law to the rate of delayed electron emissions between 2-200\,ms after the primary S2. Though this fit does not perfectly describe the data, changing the upper bound of the fit range from 200 to 700\,ms changes the value of the power law by only 5\%. The fitted power laws for position correlated 1 electron ($\gamma\,=\,-1.1$), 2 electrons ($\gamma\,=\,-1.3$), and 3-5 electrons ($\gamma\,=\,-1.4$) differ. We conclude that the 2 and 3-5 electron signals are not from pile-up of single electrons, as the fitted power law for these populations are larger than would be expected from pile-up~\cite{Kopec:2021ccm}, based on the fitted value of the 1 electron correlated population. Thus the mechanism that produces the delayed single electrons is either able to produce few-electron signals or multiple mechanisms are involved. The rate of position uncorrelated electrons drops sharply beyond the maximum drift time and then reduces more slowly in time, and is almost constant after $\mathcal{O}(10)$\,ms in the 3-5 electron population. As explained in Section~\ref{sec:prim_size}, we attribute this behavior to imperfect removal of the correlated electron emission. Indeed requiring more stringent position separation selections reduces the number of correlated electrons leaking into the uncorrelated population, flattening the observed power law. For example, changing the $\Delta \mathrm{r}$ requirement for the 1 electron position uncorrelated from 20 to 50\,cm changes the fitted power law from $\gamma\,=\,-0.72$ to $\gamma\,=\,-0.64$.

\begin{figure}[t!]
    \centering
    \includegraphics[width=0.98\columnwidth]{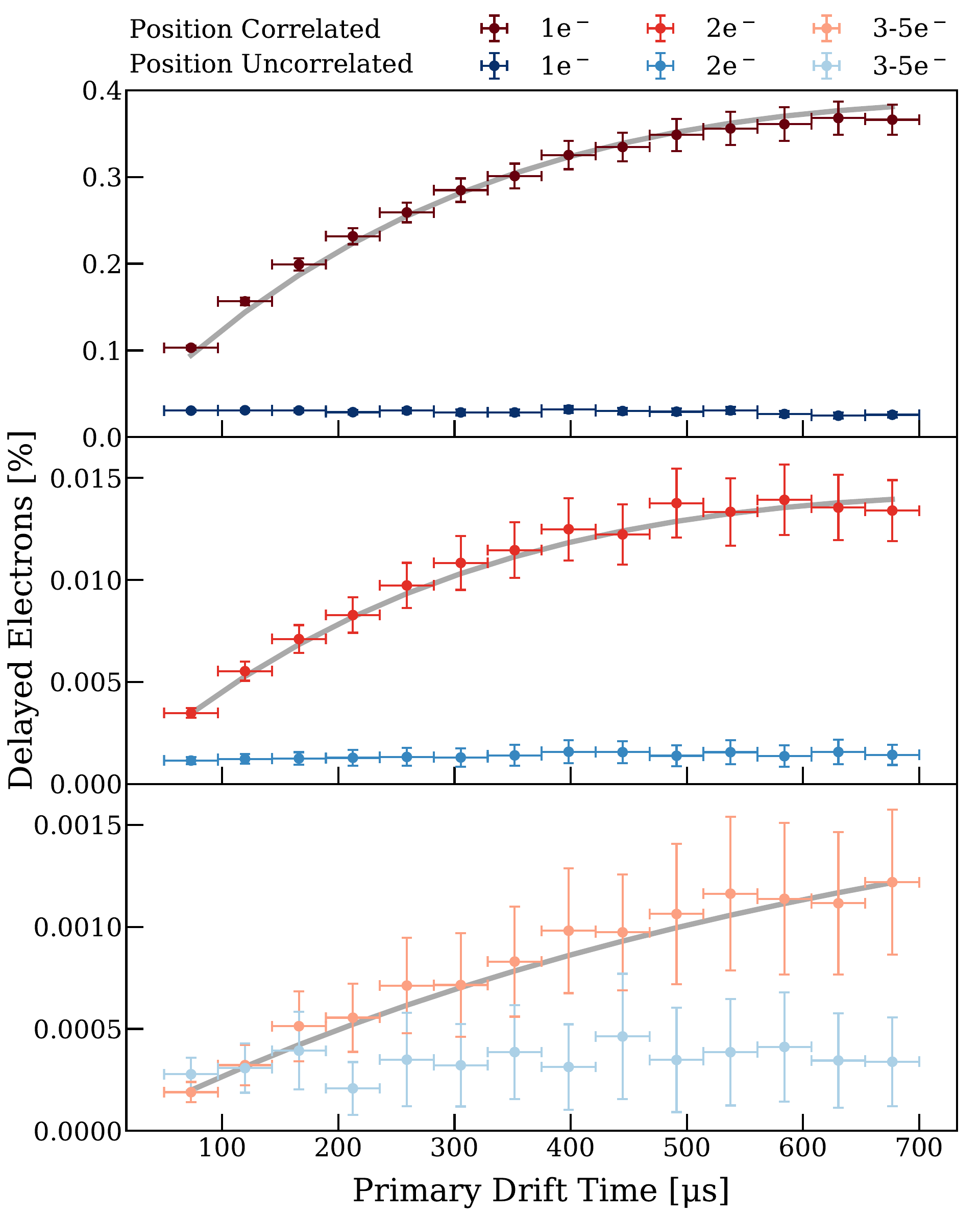}
\caption{The intensity of delayed electrons as a function of drift time within the detector between 2-200\,ms after the primary S2. The top, middle and bottom panels show the delayed electron intensity in the 1, 2 and 3-5 electron populations, respectively. Shown in gray in each panel is the expected number of delayed electrons assuming the rate at which delayed electrons are produced is constant along the full drift length and proportional to the number of electron captured on impurities.  Position correlated (uncorrelated) electrons are shown in red (blue). The position uncorrelated electrons do not show a correlation with the drift time of the primary S2, thus they are attributed to the delayed electron emission from previous interactions.}
    \label{fig:drift}
\end{figure}

\subsection{Drift Time Dependence} \label{sec:drift}
Given the observed dependence of delayed electron emissions on primary S2 size (Section~\ref{sec:prim_size}) and position in the x-y plane (Section~\ref{sec:xy_pos}), we further study any dependence on the drift time of the cloud of ionization electrons produced by the original interaction, using the continuous dataset. We split our data into 14 bins based on the drift time of the primary S2, ranging from 50\,\textmu s (just below the gate) to 700\,\textmu s (just above the cathode). The intensity of delayed electron emission between 2-200\,ms after the primary S2 is shown in Figure~\ref{fig:drift} for position correlated (uncorrelated) electrons in red (blue), subdivided into 1, 2 and 3-5 electron populations. As the size of the primary S2 is reduced due to electronegative impurities within the LXe capturing electrons along the drift track, we use the electron lifetime corrected S2 (cS2) when calculating the percentage of delayed electron emission.

As is visible in Figure~\ref{fig:drift}, the intensity of position correlated delayed electron emission increases with larger drift time in the detector. Shown in the gray line is the expected number of delayed electrons under the assumption that electrons are captured at a constant rate along the drift track. The expected number of delayed electrons is determined by accounting for the known electron lifetime in the continuous data used here. We also account for the possibility that delayed electrons that are emitted several to hundreds of millisecond after the primary S2 can be recaptured by impurities as they drift towards the liquid-gas interface. The expected number of delayed electrons are normalized to the value of the lowest drift time bin, where the impact of the secondary losses are minimized due to the shorter drift time. For the data used in Figure~\ref{fig:drift}, the electron lifetime was measured to be 660\,\textmu s. We observe good agreement between data and expectation for all 3 populations.

Contrary to correlated electrons, the intensity of position-uncorrelated electrons shows no equivalent dependence on the drift time. In Sections~\ref{sec:prim_size} and~\ref{sec:time} we do observe dependence of the uncorrelated delayed electron emission on the size of and time since the primary S2. Here, however, we do not observe an equivalent dependence as we have averaged out over the primary S2 size and integrated the number of delayed electrons observed in the 2-200\,ms window after the primary S2. This observation is consistent with the theory that delayed electron emission is a byproduct of impurities in the LXe \cite{Sorensen:2017kpl}, as longer drift times provide more opportunities for electrons to be trapped by impurities along the drift track of the electron cloud from the primary S2.  

\subsection{Extraction Field Dependence} \label{sec:efield}
The source of delayed electron emission could be electrons with insufficient kinetic energy to cross the barrier at the liquid-gas interface on their first attempt~\cite{Sorensen:2017ymt}. These electrons are trapped at the interface, but are subject to the strong extraction field in the LXe of 4.1\,kV/cm between the gate and anode electrodes in standard data taking conditions. They could consequently continue to attempt to tunnel through the barrier in the $\mathcal{O}(10)$\,ms following the primary S2. Therefore, we study the effect of the extraction field on the intensity of delayed electrons. The intensity is expressed as a percentage of delayed electrons observed from 2-200\,ms after the primary S2, relative to the number of electrons in the primary S2. We use data from the continuous dataset taken at different anode voltages. The efficiency of extracting electrons from the liquid to the gas is dependent on the  extraction field between the gate and the liquid-gas interface~\cite{Xu:2019dqb}. 

\begin{figure}[t!]
    \centering
    \includegraphics[width=0.92\columnwidth]{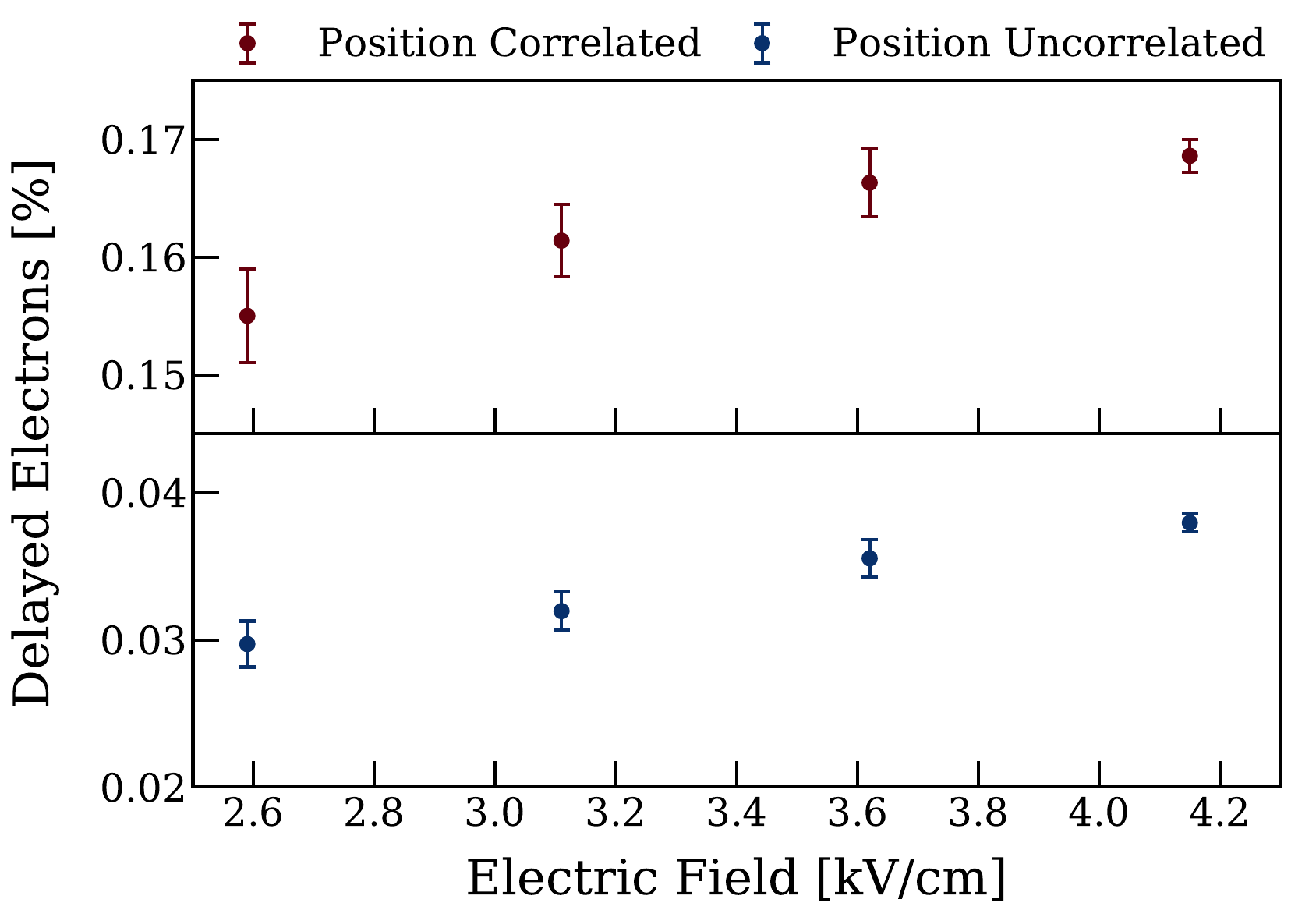}
    \caption{The intensity of delayed single electrons as a function of the extraction field in the liquid xenon. Delayed electrons emitted between 2-200\,ms after the primary S2 are shown as a percentage of electrons in the primary S2, for position correlated (uncorrelated) single electrons, red points in top figure (blue points in bottom figure). The intensities of the delayed electron emission for 2 and 3-5 electrons are not shown, but found to also be weakly correlated with the extraction field.}
    \label{fig:efield}
\end{figure}

If delayed electron emission in XENON1T is primarily due to electrons trapped at the liquid-gas interface, then we would expect to observe a dependence on the intensity thereof with changing extraction field. As the number of delayed electrons should be proportional to the number of electrons that reached the liquid-gas interface, we use the uncorrected size of the primary S2. We observe a $\leq$0.02\% absolute (15\% relative) variation in the intensity of position correlated delayed electrons as we vary the extraction field, as shown in Figure~\ref{fig:efield}. The changes in delayed electron emission with variations in extraction efficiency are however less significant than that observed for drift times as discussed in Section~\ref{sec:drift}. Thus, we conclude that it is unlikely that the delayed electron rate is primarily affected by the extraction field in the 2-200\,ms window after an interaction, though we cannot rule out that this is a secondary effect. The small variations with extraction field, in comparison to other effects studied in this work, seems to disfavor trapped electrons being the cause of the delayed electrons in XENON1T.

\subsection{Dependence on Date and Purity} \label{purity}
\begin{figure}[t!]
    \centering
    \includegraphics[width=0.98\columnwidth]{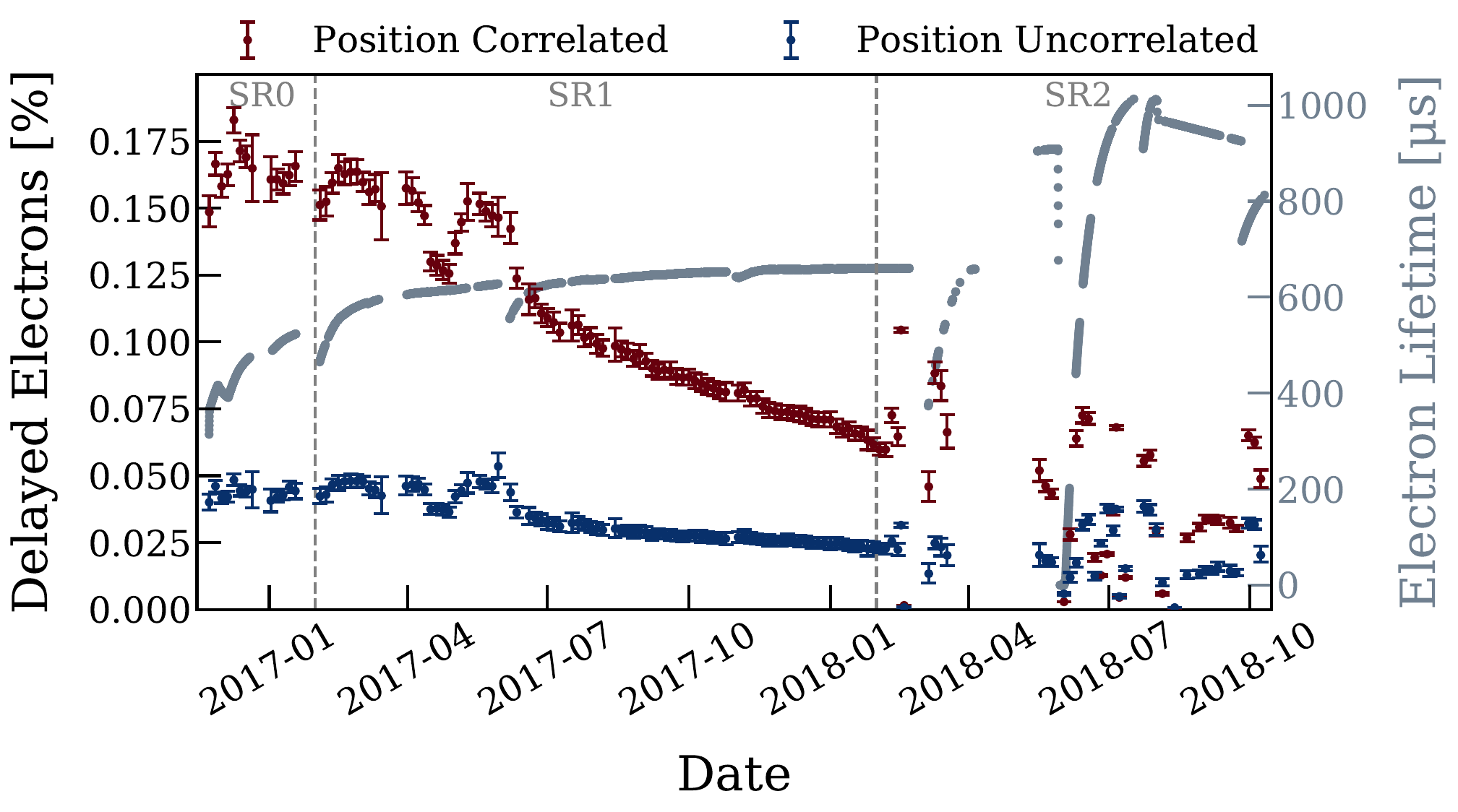}
    \caption{The intensity of delayed electron emissions between 2-200\,ms after the primary S2 as a function of date. Data are shown for position correlated (uncorrelated) delayed single electron emission red (blue). Also shown in gray are the measured electron lifetime values in each time period. The end of the first and second science runs are indicated by gray dashed lines.}
    \label{fig:date}
\end{figure}

Due to the strong dependence on the drift time of the primary interaction, described in Section~\ref{sec:drift}, impurities seem to be the dominant origin of delayed electron emission. Throughout XENON1T's lifespan, the LXe target was constantly purified by means of gaseous recirculation passing through gas purifiers. In the first two science runs (SR0, SR1) this constant purification resulted in continually improving LXe purity within the detector, reaching a maximum measured electron lifetime of $\sim$660\,\textmu s. During the last year of operation (SR2), several R\&D initiatives (for instance~\cite{XENON:2021fkt}) allowed for greatly improved purity conditions, reaching electron lifetime values in excess of 1\,ms as shown in the gray data points in Figure~\ref{fig:date}. Also shown in Figure~\ref{fig:date} in red (blue) are the data points of the measured intensity of position correlated (uncorrelated) delayed single electron emission as a function of date using the triggered dataset. The behavior for 2 and 3-5 electrons is observed to be the same as for single electrons. The intensity of the delayed emission is given as a percentage of electrons in the primary cS2. We observe that the intensity of position correlated electron emissions decreases gradually throughout the first two science runs with increasing LXe purity. The intensity of delayed electron emission is more variable in SR2 due to frequent interventions on the detector, but the overall decreasing trend in the intensity of delayed electron emission with time is still observed. The intensity of position uncorrelated delayed electron emission displays a much smaller decrease throughout the time period during which XENON1T collected data.

\begin{figure}[t!]
    \centering
    \includegraphics[width=0.98\columnwidth]{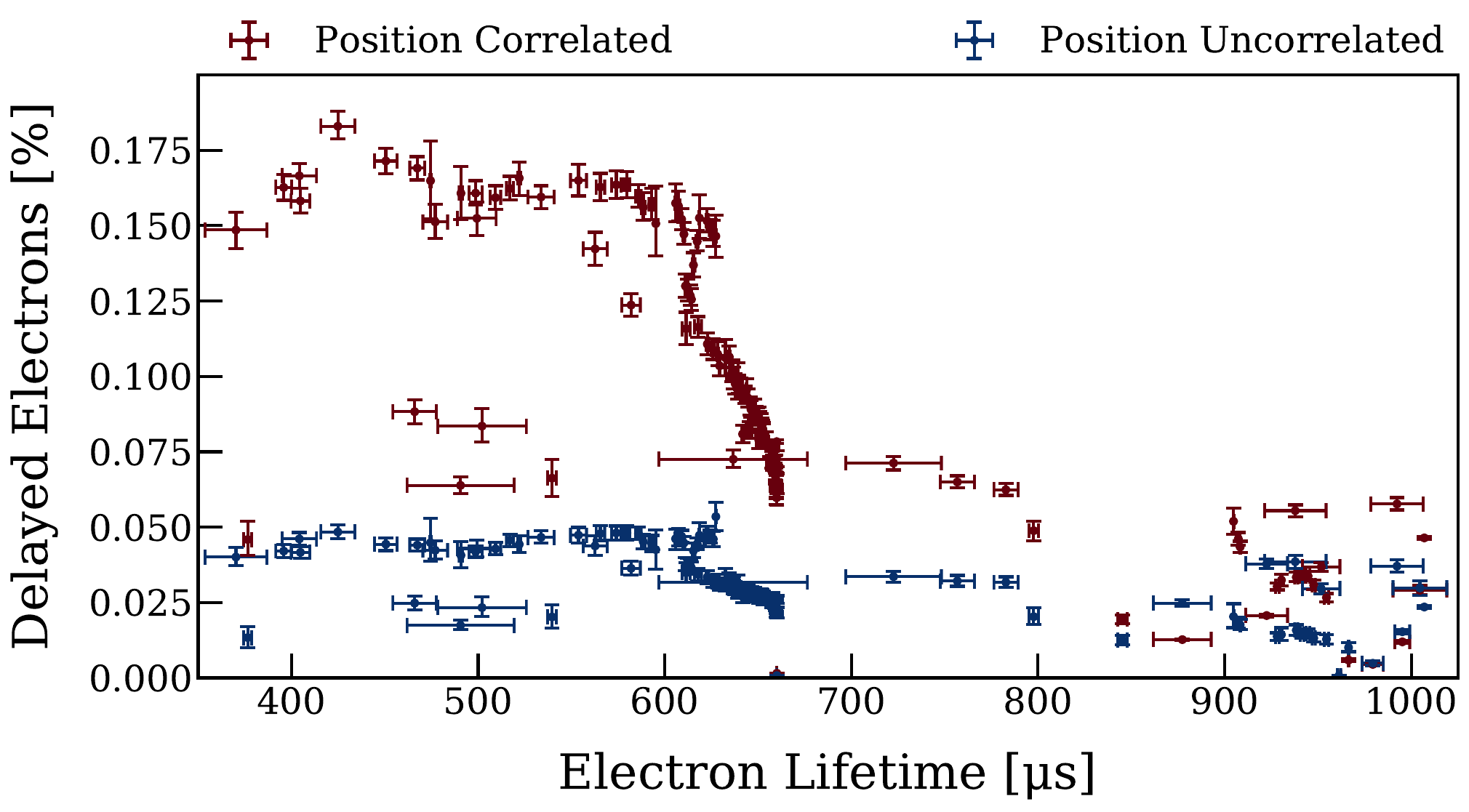}
    \caption{The intensity of delayed electron emissions between 2-200\,ms after the primary S2 as a function of electron lifetime. Data are grouped into four-day periods and shown as a function of the average electron lifetime measurement in each time period. Position correlated (uncorrelated) delayed single electron emission is shown in red (blue).}
    \label{fig:purity}
\end{figure}

We additionally plot the intensity in Figure~\ref{fig:date} as a function of the average electron lifetime measured in each period using the triggered dataset, shown in Figure~\ref{fig:purity}. We observe that data taken with equally pure LXe, but separated in time, do not produce the same intensity of delayed electron emission. Specifically, data collected during SR2 show lower intensity of position correlated single electron emission while obtaining the highest electron lifetime. In fact, the greatest change in LXe purity in XENON1T, occurring in July 2018, shows very little change in the intensity of the delayed electron emission. We conclude that both the correlated and uncorrelated electrons display a weak dependence on the measured electron lifetime. Note, that the apparent rapid decrease in the fraction of delayed electrons extracted between 600 and 660\,\textmu s is an artifact of our electron lifetime improving in time as shown by the gray data points in Figure~\ref{fig:date}.

\subsection{Summary of Background Characterization}
From the observed intensity of delayed electron emission in XENON1T, we conclude that this instrumental background is observed only for interactions in the drift region of the detector (\ref{sec:primary}). The delayed electron emission is observed to occur in close proximity to the most recent interaction (\ref{sec:xy_pos}). In addition, the intensity of the emission is found to increase with the number of electrons observed from the original interaction (\ref{sec:prim_size}) and to be observed over $\mathcal{O}$(100)\,ms (\ref{sec:time}) after the interaction. We limited ourselves to studying the delayed electron emissions between 2-200\,ms after a primary S2. Beyond 200\,ms the data is insufficient to study the behavior of delayed electron emission for each of the 1, 2 and 3-5 electron populations separately.

We conclude that delayed electron emission observed in XENON1T is not primarily a product of imperfect extraction of electrons at the liquid-gas interface. This is because the observed correlation between the extraction field and the intensity of the emission (\ref{sec:efield}) is small relative to other effects studied in this work. Rather, the increased emission observed for deeper interactions (\ref{sec:drift}) leads us to conclude that some impurity within the LXe itself is responsible for the bulk of the delayed emission. However, we do not observe consistent intensity of delayed electron emission for periods separated in time when the detector had similar electron lifetime (\ref{purity}) conditions. Rather  we observe a continual decrease of the delayed electron emission in time during the operational lifetime of XENON1T. In XENON1T more than half of the electrons from the deepest interactions were captured by impurities. Impurities that impact the electron lifetime, such as O$_2$, will release trapped electrons within hundreds of microseconds of the primary S2 due to photoionization in the detector. In contrast, the fraction of electrons observed as delayed electron emission represents less than 1\% of the drifting electrons. Therefore, we believe that if an impurity species within the LXe is responsible for the delayed electron emission, this specific species of impurity does not strongly affect the calculated electron lifetime, and thus the assumed purity, of XENON1T. Since these delayed electron emissions are observed hundreds of milliseconds after the primary S2, we speculate that an impurity other than O$_2$ may be trapping these electrons.

%% file: sec_4_cuts.tex
\section{Dark Matter search selections} \label{dm_cuts}

We can use the results from Section~\ref{characterization} to perform a low-background search for DM by developing selections based on the bulk delayed electron population to suppress the time and position correlated single- and few-electron events. This allows us to push our analysis threshold down to a single detected electron, thereby extending our DM search to lower masses than the previous XENON1T S2-only analysis~\cite{XENON:2019xxb}. We will conservatively assume that all surviving few-electrons are due to DM as we will set only upper limits on various DM models.

\begin{figure}[t!]
    \centering
    \includegraphics[width=0.98\columnwidth]{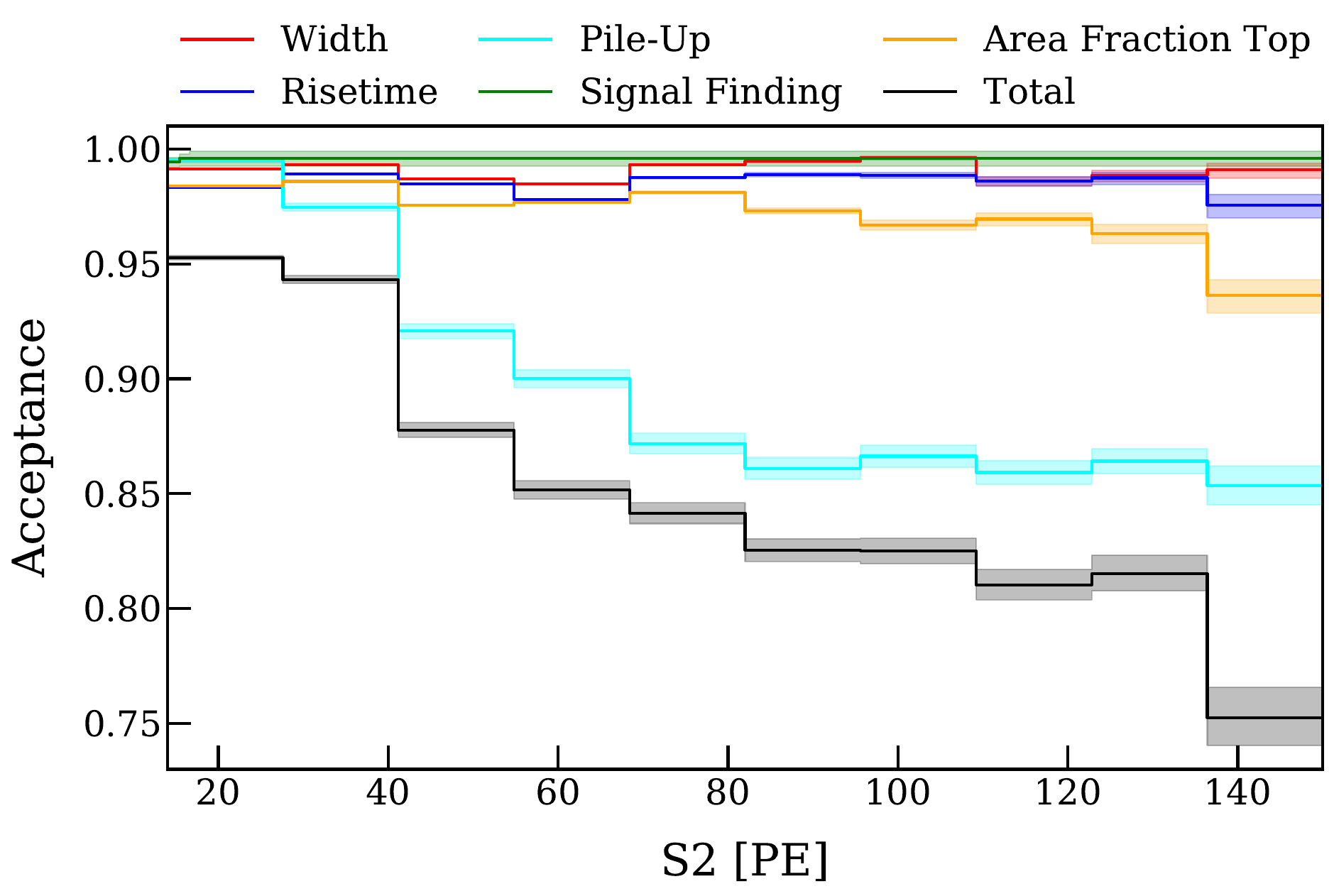}
    \caption{The acceptance of signal selection criteria of S2s in our ROI. Acceptances are calculated for the case where the given selection is applied last. The most impactful selection is the ``pile-up'' selection shown in cyan, which removes events where two spatially separated signals are reconstructed as a single signal. Other selections are constructed to have high acceptance within our ROI and primarily exclude misidentified S1s. The shaded regions indicate the 1$\sigma$ confidence interval of each selection.}
    \label{fig:cut_efficiency}
\end{figure}

We apply two broad categories of event selections. The first category are selections based on the dependence between the primary S2 and the delayed electron population, used to select volumes and times in the detector when the rate of delayed electron emission is minimal. Due to its higher livetime, we select only the continuous data taken during February 2018, with an extraction field of 4.1\,kV/cm, for setting limits on the models described in Section~\ref{sec:dm_models}. We use one day of this continuous data as a training dataset in order to identify windows of time during which we do not expect any correlated electron backgrounds, by accounting for the observed power law of the delayed electron emission. The selection of small S2 signals into 1, 2 and 3-5 electron populations is the same as that used in Section~\ref{characterization}. We split the data into 1 electron (14-42\,PE), 2 electron (42-70\,PE), and 3-5 electron (70-150\,PE) populations, where the bins are centered at multiples of the secondary scintillation gain of (28.8$\pm$0.1)\,PE. For each electron population, we fit a power law to the training dataset to create a model of the delayed few-electron population. The optimal selection on delay time from the most recent primary S2 is then determined by maximizing our exposure while minimizing the projected delayed few-electron population. As the intensity of delayed electron emissions is also found to be dependent on the size of the primary S2, the delay time for each population is determined in five different primary S2 size bins. The optimized delay time selections are dependent on both the size of the most recent primary S2 and the size of the few-electron signals. We require longer delay times for more energetic primary interactions and require shorter delay times for larger few-electron signals. We exclude primary S2 and delayed electron population bins for which no optimal solution is found, either due to a lack of statistics or high rate of delayed electron emission. 

The pre-trigger selection, described in Section~\ref{data_selection}, is applied to remove S1s and any associated electron emission due to photoionization that occur before the primary S2. This reduces the exposure following each primary S2 by 1\,ms in the continuous dataset. We limit ourselves to considering only time regions within the detector where the expected duration and intensity of delayed electrons emissions is well understood. Therefore we require the primary S2 event to be well reconstructed and to have occurred within the active LXe region between the gate and cathode; this reduces our overall exposure by $\sim$33\%. In addition, we exclude data from any time period in which the data acquisition system returned a \textit{busy} signal~\cite{XENON:2019bth}, further reducing the livetime by 5.4\%.

\begin{figure}[t!]
    \centering
    \includegraphics[width=0.98\columnwidth]{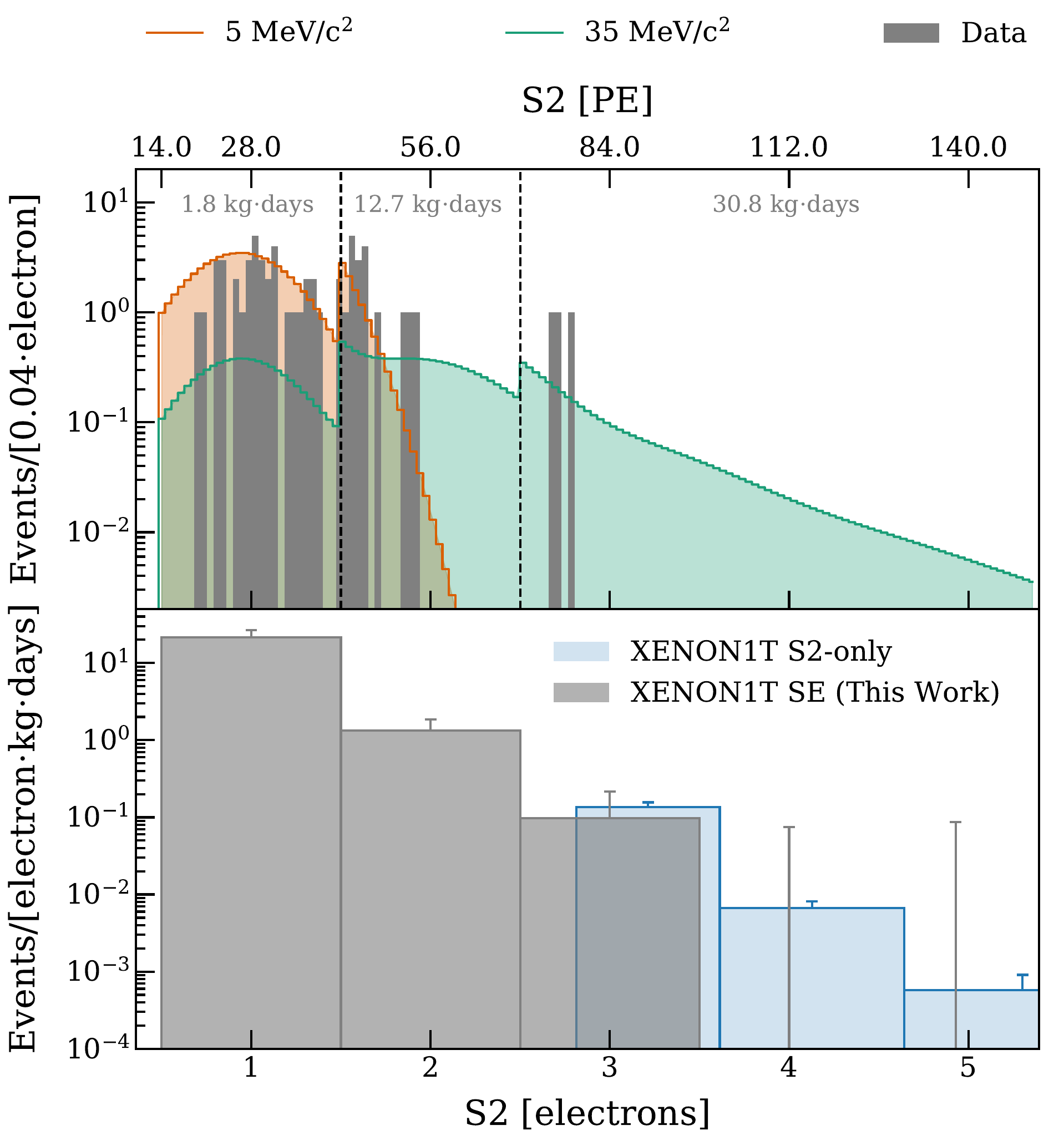}
    \caption{The distribution of events that pass all selections are shown in gray in the top panel. The boundaries of the 1, 2 and 3-5 electron search regions and the associated exposures are also indicated. The expected signal produced through DM-electron scattering via a heavy mediator for DM with mass 5 (35) MeV/c$^2$ is shown in orange (green). The final event rate, in 1 electron (28 PE) wide bins, are shown in the bottom panel for this work in gray. The event rates below 150 PE from~\cite{XENON:2019xxb} are shown in blue, which relied on a larger value of the scintillation gain due to after-pulsing in PMTs positively biasing larger S2 signals. Bars indicate the measured event rate and error bars indicate Poisson 90\% confidence level upper limit on the rates.}
    \label{fig:spectrum}
\end{figure}{}

The second category consists of criteria applied to small S2 signals observed in the selected regions in time and space. In contrast to Section~\ref{characterization}, we select small S2 signals in the full range 14-150\,PE. Our signal finding efficiency is $>$~99\% across the entire region. The fraction of light seen by the top PMTs in XENON1T is on average 63\% for S2s, with considerable energy-dependent spread especially at the lower energy threshold for this analysis. We remove events for which this fraction lies above the 99.5 percentile (indicative of events produced in the GXe phase) or below the 0.5 percentile (indicative of mis-classified S1s). Width and risetime selections, with 98\% acceptances, are applied, to further remove misidentified S1s. We remove events where two or more spatially separated few-electrons are reconstructed as a combined signal (``Pile-Up'' in Figure~\ref{fig:cut_efficiency}) by comparing the hit pattern on the top PMT array to simulated waveforms. The acceptance of these selections and the XENON1T signal finding efficiency are shown in Figure~\ref{fig:cut_efficiency}. 

Events close to the TPC walls, located at $R = 47.9$\,cm, experience charge loss, suffer from increased position reconstruction uncertainty and are affected by the drift field non-uniformity near the walls~\cite{XENON:2019izt}. We therefore remove all few-electron events with R$^2$ $\geq$700\,\textrm{cm}$^2$ (R $\geq$ 26.5\,\textrm{cm}) as was done for S2s $\leq$400\,PE in Ref.~\cite{XENON:2019xxb}. As the intensity of delayed electron emission is enhanced near the location of the primary S2 (Section~\ref{sec:xy_pos}), we also remove all few-electron signals that occur within 20\,cm of the preceding primary S2. 

After applying all selections and unblinding the data, the final exposure in the [1, 2, 3-5] electrons bins is [1.76, 12.7, 30.8]\,kg$\times$days. The final event rates are shown in Figure~\ref{fig:spectrum} along with the rates from \cite{XENON:2019xxb} with an exposure of 22\,tonne$\times$days.

%% file: sec_5_detector.tex
\section{Detector Response}\label{sec:detector_models}

Ionization electrons can be produced through light DM particles scattering off the electron cloud of a xenon atom~\cite{Essig:2011nj}, resulting in ionization of nearby atoms as the electron slows down in the LXe medium. These ionization electrons are drifted to the liquid-gas interface and are extracted into the GXe, where they emit secondary scintillation light with the same spectrum of the few-electron signals examined in Section~\ref{characterization}. In order to evaluate the expected rate of events from a specific recoil spectrum $dR/dE_\mathrm{r}$, we use the following model, accounting for the detector response and reconstruction effects:

\begin{equation}
\label{eq:detector_response}
\begin{split}
R(S2, z) & =  \varepsilon(S2)\times\sum_{n_\mathrm{e}=1}^{\infty}\sum_{k=0}^{N}\int dE_\mathrm{r}\frac{dR}{dE_\mathrm{r}}\textrm{Binom}(k|N;p_{\mathrm{qe}})\\
			&\times \textrm{Binom}\Big(n_\mathrm{e}|k;\epsilon_{\mathrm{ext}}\times e^{(-z/\tau_\mathrm{e}\times v_\mathrm{d})}\Big)\\&\times\textrm{Normal}(S2|\mu_{\mathrm{S2}},\sigma_{\mathrm{S2}})
\end{split}
\end{equation}

\noindent where $\varepsilon(S2)$ is the combined selection efficiency shown in Figure~\ref{fig:cut_efficiency} and $n_\mathrm{e}$ is the number of electrons extracted into the GXe. Here, integer $N=E_\mathrm{r} Q_\mathrm{y}$ ionization or excitation quanta will be created from an ER of energy $E_\mathrm{r}$, where $Q_\mathrm{y}$ is the energy-dependent charge yield. From these quanta, $k$ will be observed as electrons, with probability $p_{\mathrm{qe}}=(1-\langle r\rangle)/(1+\langle N_{\mathrm{ex}}/N_{\mathrm{i}}\rangle)$, a function of the ER mean recombination fraction $\langle r\rangle$ and of the exciton-to-ion ratio $\langle N_{\mathrm{ex}}/N_{\mathrm{i}}\rangle$. The depth at which an ionization electron is created is given by $z$, while $\tau_\mathrm{e}$ is the electron lifetime and $v_\mathrm{d}$ is the electron drift velocity. The average value of the extraction efficiency is denoted by $\epsilon_{\mathrm{ext}}$. Finally, the secondary scintillation light produced by an electron extracted into the GXe is modeled as a Gaussian with mean $\mu=n_\mathrm{e} G(1+\delta_{\mathrm{S2}})$ and $\sigma=\sqrt{n_\mathrm{e} \Delta G^2+ (\mu\Delta\delta_{\mathrm{S2}})^2}$. Here $G$ and $\Delta G$ are the secondary scintillation gain and its spread. $\delta_{\mathrm{S2}}$ and $\Delta\delta_{\mathrm{S2}}$ are the software reconstruction bias and its spread. 

The value of the aforementioned quantities are taken to be those reported in~\cite{XENON:2019izt} apart from the value of $G$ which is taken to be (28.8$\pm$0.1)\,PE, and $\Delta G$ which is (7.13$\pm$0.04)\,PE. This value is lower than in~\cite{XENON:2019izt} as the small, few-electron signals studied here are not affected by PMT after-pulsing which tends to positively bias larger S2 signals. We also account for the software bias between the number of reconstructed photoelectrons and the expected number of detected photoelectrons. This reconstruction bias, $\delta_{\mathrm{S2}}$, and its spread, $\Delta\delta_{\mathrm{S2}}$, are estimated as functions of S2 via waveform simulations~\cite{XENON:2019ykp}.

The charge yield is defined as $Q_\mathrm{y}=p_{\mathrm{qe}}/W$. We assume W=13.8\,eV in this work, though recent measurements have suggested a value of 11.5\,eV ~\cite{EXO-200:2019bbx, Baudis:2021dsq} for the average energy required to produce an excitation quanta in xenon. Using the lower value would increase the expected signal rate in our ROI by a maximum value of 4\%. As the higher value is consistent with previously published limits on light DM and results in more conservative limits, we do not update the value of W in this work. 

\begin{figure}[t!]
    \centering
    \includegraphics[width=0.98\columnwidth]{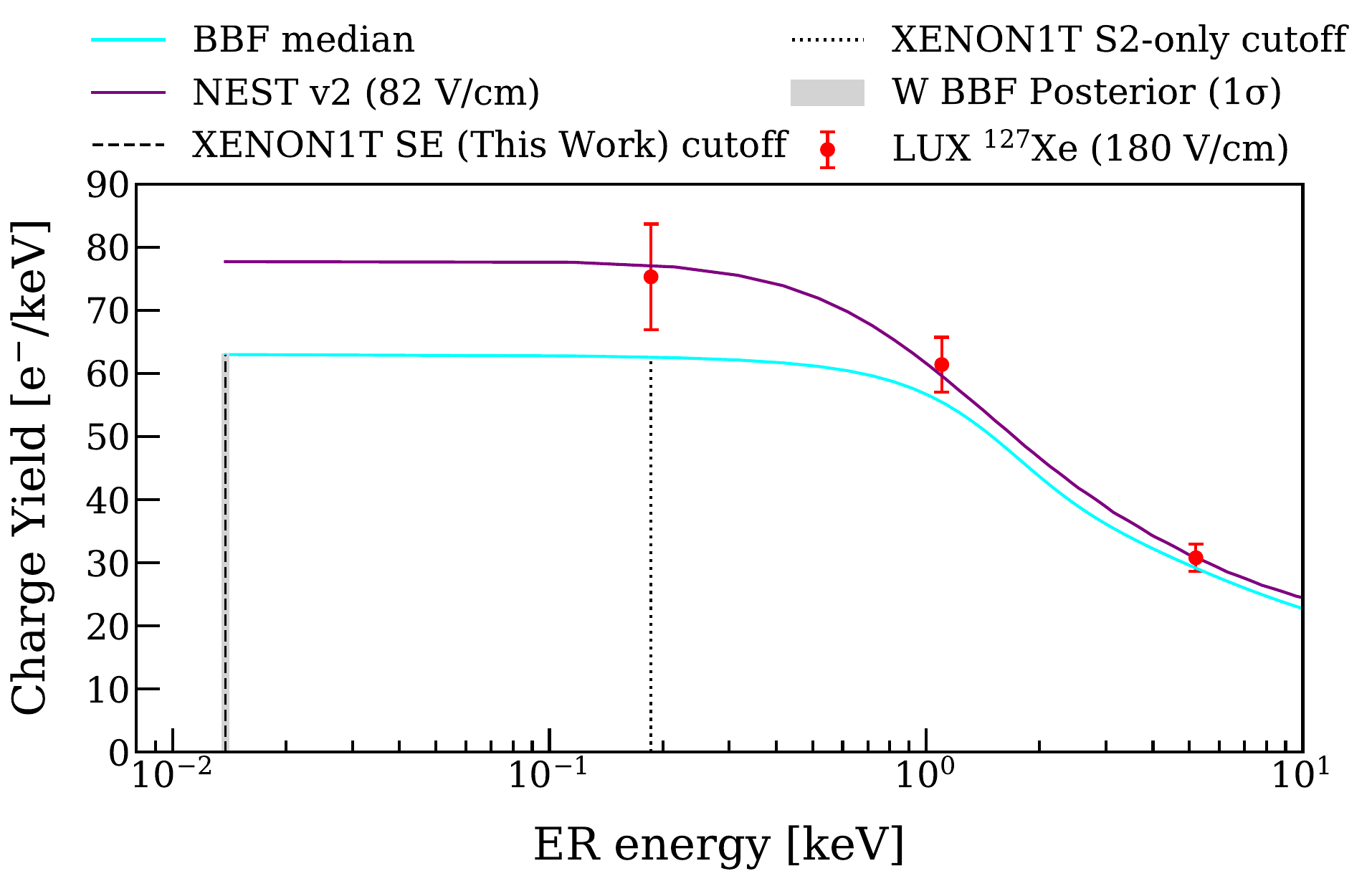}
    \caption{Charge yield for LXe as a function of ER energy. Shown are the BBF median (cyan) and posterior (gray band) from \cite{XENON:2019izt}, the best fit of NEST~v2~\cite{szydagis_m_2018_1314669} (purple) as well as measurements from LUX~\cite{LUX:2017ojt} (red points). Indicated by the dashed line is the energy cutoff used in this work. The previous XENON1T S2-only analysis~\cite{XENON:2019xxb} applied an energy cutoff at 0.186\,keV indicated by the dotted line.}
    \label{fig:charge_yield}
\end{figure}

In this analysis we use the modified Thomas-Imel box model~\cite{Thomas:1987zz, XENON:2018dbl} to describe the recombination fraction $\langle r\rangle$. For the nominal values of the model, we use the median of the posterior obtained from the best fit of the model to the XENON1T ER calibration data using a Bayesian simultaneous fit (BBF) framework~\cite{XENON:2019izt}. The ER low energy calibration was performed using a $^{220}\textrm{Rn}$ source, exploiting the $\beta$-decay of its $^{212}\textrm{Pb}$ progeny to the ground state of $^{212}\textrm{Bi}$~\cite{XENON:2017iyp}. The detection efficiency of $^{212}\textrm{Pb}$ decays drops off below 1.6\,$\mathrm{keV}$ ER equivalent if one relies on detection of both the S1 and S2 signal. Therefore, in XENON1T’s S2-only search~\cite{XENON:2019xxb}, a lower cutoff was set for the ER charge yield. This point corresponds to the lowest-energy absolute calibration of the charge yield performed in~\cite{LUX:2017ojt} using electron capture events occurring from the N-shell of $^{127}\textrm{Xe}$. Our region of interest (ROI) corresponds to deposited energies in the sub-keV region below this threshold. However, relevant results have been published using an extrapolation of the LXe response to ERs assuming only a theoretical understanding of the ionization process~\cite{Essig:2012yx, Essig:2017kqs, Bloch:2016sjj, PandaX-II:2021nsg}. Here we extrapolate the best fit of the BBF Thomas-Imel model to the SR1 $^{220}\textrm{Rn}$ calibration data, as shown in Figure~\ref{fig:charge_yield}. The value of the SR1 derived charge yield is systematically lower than that derived from the Noble Element Simulation Technique (NEST~v2) model when extrapolated into our ROI using the SR1 drift field of 82\,kV/cm~\cite{szydagis_m_2018_1314669}. In our ROI, the ER charge yield derived from the NEST~v2 model is almost independent of the electric field. Our detector response model can thus be considered to be conservative. We do not proceed with an extension of the S2-only analysis for a SI WIMP-nucleon interaction although suggestions do exist about various extrapolations below the lower-energy NR charge yield measurement~\cite{Wang:2016obw, Essig:2018tss}.

{}

As we cannot reconstruct the depth of an interaction in our ROI without the observation of an S1, the data are compared to the distribution of expected events only through the projection of the 2D map (Eq.~\ref{eq:detector_response}) onto the S2-space. The recoil rate of a given signal model is considered to be uniform in z, thus the only dependence on the depth originates from the electron drift term in Eq.~\ref{eq:detector_response}. We also take into account the discontinuous exposure in the S2-space, shown in Figure~\ref{fig:spectrum}.

% Description of other backgrounds
Previous S2-only studies in XENON1T~\cite{XENON:2019xxb} have considered additional background sources in the form of coherent elastic nuclear scattering of solar neutrinos (CE$\nu$NS), neutrino-electron interactions, and ER backgrounds from detector components and intrinsic backgrounds. Solar neutrinos scatter off the nuclei of xenon atoms, where the dominant contribution to CE$\nu$NS is from the flux of $^{8}$B neutrinos~\cite{Billard2014}. These interactions produce only a few ionization electrons and are observed as lone-S2 events. The resultant spectrum is remarkably similar to the spectrum of a WIMP with $m=6$\,GeV/c$^2$ and $\sigma_{SI}=$\,\num{4.7e-45}\,cm$^2$~\cite{Billard2014, COHERENT:2017ipa, Serenelli:2011py}. In the case of neutrino-electron interactions, we consider the dominant sources of solar neutrinos, namely \emph{pp} fusion and electron capture by $^{7}$Be. We account for the electron binding energies as in Ref.~\cite{Chen:2016eab} and find that the differential recoil energy spectrum is maximal at energies well above our ROI. Finally, we also consider the ER background at low energy. This background is dominated by the low energy tail of the $\beta$-decay of $^{214}\text{Pb}$ to the ground state of ${^{214}\text{Bi}}$. Smaller contributions from $^{85}$Kr and the material background generated by Compton scattering of gamma rays reaching the internal volume are also present. The rates for these three ER background sources are taken from Ref.~\cite{XENON:2020rca} and are modeled together as a single background. The combined model is extended in our ROI as a flat rate of \num{1.42e-4}\,events/(kg$\times$day$\times$keV). 

Using the detector response described above, we determine the expected event rates from CE$\nu$NS events, neutrino-electron interactions and the flat ER background. These are compared to the delayed electron background shown in Figure~\ref{fig:spectrum}. We find that the contribution from all three known physical backgrounds are smaller by several orders of magnitude, thus we conservatively neglect them in the limit setting procedure. Thus, for an S2-only analysis with a ROI corresponding to 1-5 electrons, delayed electron emission is the dominant background. As our understanding of this background is empirical and not derived from a physical model of a known source, we cannot perform a background subtraction.

%% file: sec_6_models.tex
\section{Limits on Dark Matter Models}\label{sec:dm_models}

We set constraints on a number of light DM models with sub-GeV masses. In these models, interactions between DM and electrons are mediated by a dark-sector gauge boson coupling with charged standard model particles via kinetic mixing with a photon~\cite{Holdom:1985ag}. We use the optimum-interval method from Ref.~\cite{Yellin:2002xd} to compare signal models to our data in the S2 region from 150\,PE (5 electrons) down to a single detected electron, shown in Figure~\ref{fig:spectrum}. The expected detector response from a specific signal model is obtained as described in Section~\ref{sec:detector_models}. We compute an aggregate uncertainty on the signal expectations in our ROI, accounting for a $\sim$5\% uncertainty on electron lifetime and $\sim$2.5\% uncertainty on the S2 gain. 

Limits have previously been reported for all the models we probe in this work. Figure~\ref{fig:spectrum} illustrates the signal expectation for DM-electron scattering via a heavy mediator, with DM mass of 5\,MeV/c$^2$ (35\,MeV/c$^2$) and cross-section of \num{4.6e-34}\,cm$^2$ (\num{3.7e-38}\,cm$^2$) in orange (green). The exact signal expectation can be influenced by detector conditions such as the electron lifetime, selection and trigger efficiencies, and reconstruction biases. In addition, the assumptions made about the microphysics response of LXe, such as the assumed charge yield and W value, may not be consistent as more recent measurements have become available. Therefore, where results have been calculated indirectly from previous experimental results, and where assumptions have been made on the exact detector response, we choose to represent the limit as gray lines. Those results should be used with caution, as they may not be comparable to this work.

\paragraph{\textbf{DM-electron Interactions}} We consider the case of DM-electron scattering, in which a fermion or scalar boson DM candidate scatters off an electron bound in a xenon atom. We follow the approach laid out in Ref.~\cite{Essig:2015cda}, as described in Appendix~\ref{sec:dm_e_scattering}. We treat the target xenon atoms as isolated, resulting in assumed binding energies (12.1\,eV~\cite{Essig:2015cda}) larger than the true binding energy due to the electronic band structure of LXe (9.2\,eV~\cite{schmidt91}). Our estimation of the ionization rate can thus be considered to be  conservative~\cite{Steinberger1973}. The interaction cross section is dependent on the DM form factor, for which we consider two benchmark models:

\begin{itemize}
  \item $F_{\mathrm{DM}}(q)=1$, where the scattering can be approximated as a point-like interaction, for example resulting from a heavy vector mediator exchange. 

  \item $F_{\mathrm{DM}}(q)=(\frac{\alpha m_\mathrm{e}}{q})^2$, where the interaction occurs via the exchange of an ultra-light vector mediator.
\end{itemize}

\begin{figure}[t!]
    \centering
    \includegraphics[width=0.98\columnwidth]{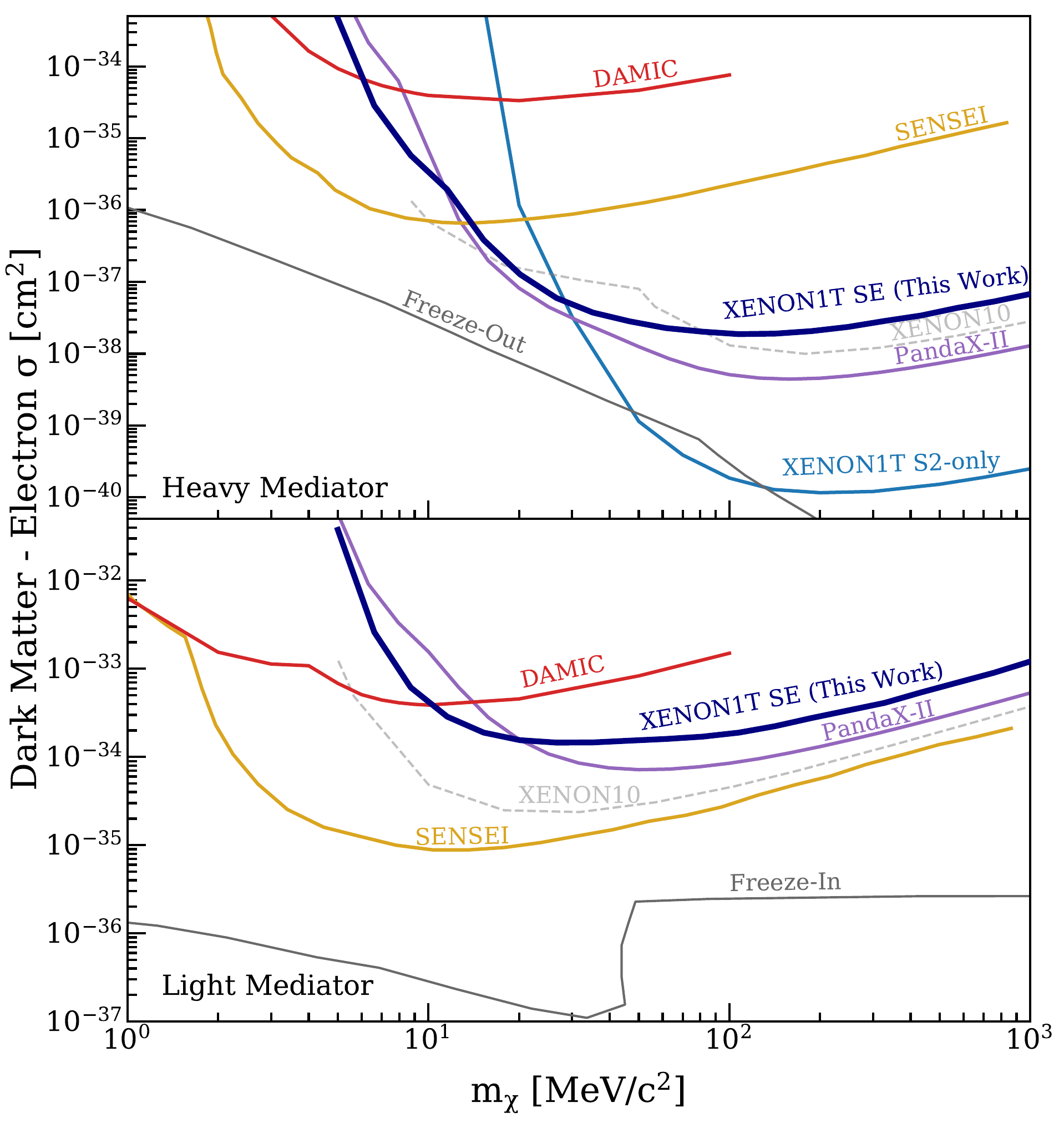}
    \caption{The 90\% confidence level upper limits on DM-electron scattering (dark blue) via a heavy mediator (top) and a light mediator (bottom), as function of DM mass $m_{\chi}$. For comparison, we show experimental results (solid) from XENON1T S2-only~\cite{XENON:2019xxb} (light blue), PandaX-II~\cite{PandaX-II:2021nsg} (purple), SENSEI~\cite{SENSEI:2020dpa} (gold) and DAMIC~\cite{DAMIC:2019dcn} (red), alongside limits calculated (gray) in Ref.~\cite{Essig:2017kqs} using data from XENON10~\cite{XENON10:2011prx} (dashed). Additionally shown is the relic abundance from freeze-out (top) and freeze-in (bottom)~\cite{Essig:2015cda} (dark gray).}
    \label{fig:dm-electron}
\end{figure}

The 90\% confidence upper limits for DM-electron scattering are shown in Figure~\ref{fig:dm-electron}, where direct experimental results are shown in solid lines, and limits recast from experimental data with inferred detector response models are shown in gray lines. By lowering XENON1T's S2 threshold to include single electron signals, we are able to probe DM-electron scattering via the light mediator, which was not done in Ref.~\cite{XENON:2019xxb}. The ability of XENON1T to set strong limits on DM signals that are dominated by few-electron signals is degraded due to the fact that the electron lifetime and maximum electron drift time in XENON1T are both $\mathcal{O}(1)$ms. This occurs due to both the large background in the 14-42\,PE region (1 electron) from impurities, as well as the fact that electron loss from larger DM signals ($\geq$ 2 electrons) deeper in the detector would result in the signal spectrum being strongly peaked in the 1 electron region. 

\begin{figure}[t!]
    \centering
    \includegraphics[width=0.98\columnwidth]{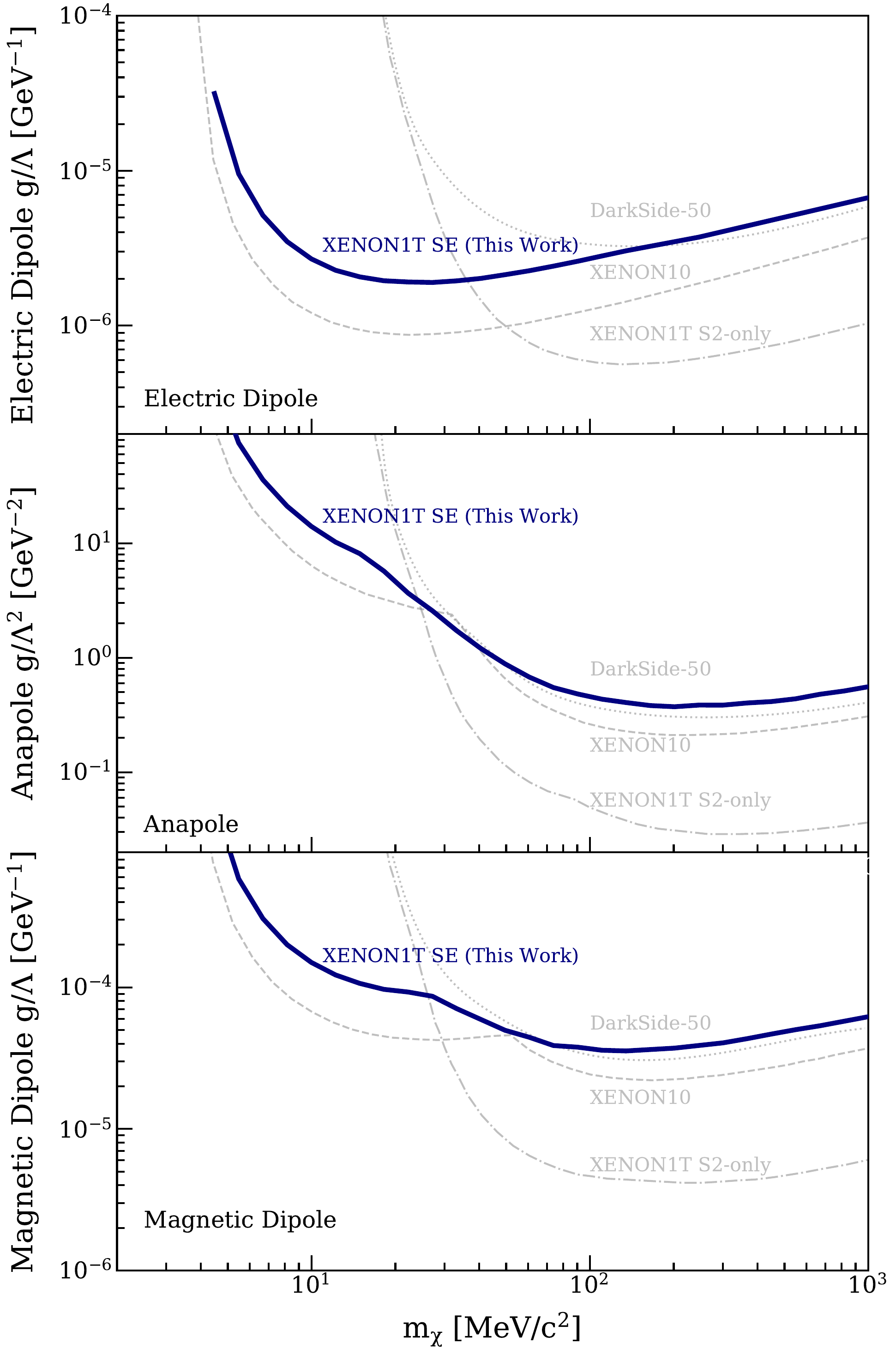}
    \caption{The 90\% confidence level upper limits on electric dipole (top), anapole (middle), and magnetic dipole (bottom) interactions (dark blue) as function of DM mass $m_{\chi}$. For comparison, we show limits calculated (gray) in Ref.~\cite{Catena:2019gfa} using data from XENON10~\cite{XENON10:2011prx} (dashed), XENON1T S2-only~\cite{XENON:2019xxb} (dot-dashed) and DarkSide-50~\cite{DarkSide:2018ppu} (dotted).}
    \label{fig:eft}
\end{figure}

Implied in the treatment of the DM-electron scattering cross section described in detail in Appendix~\ref{sec:dm_e_scattering}, is the assumption that the scattering amplitude is dependent only on the transferred momentum. We use a non-relativistic effective theory~\cite{Catena:2019gfa} to derive the most general form of this amplitude, for which we investigate the effective coupling constants on three models of DM-electron interactions, namely anapole, magnetic dipole, and electric dipole interactions. The limits are set on the ratio $g/\Lambda^2$ for anapole and $g/\Lambda$ for magnetic dipole and electric dipole interactions, where g is the dimensionless coupling constant and $\Lambda$ is the energy scale at which the corresponding interaction is generated. These limits are shown in Figure~\ref{fig:eft}, and represent the first direct limits from experimental results on these operators. 

\begin{figure}[t!]
    \centering
    \includegraphics[width=0.98\columnwidth]{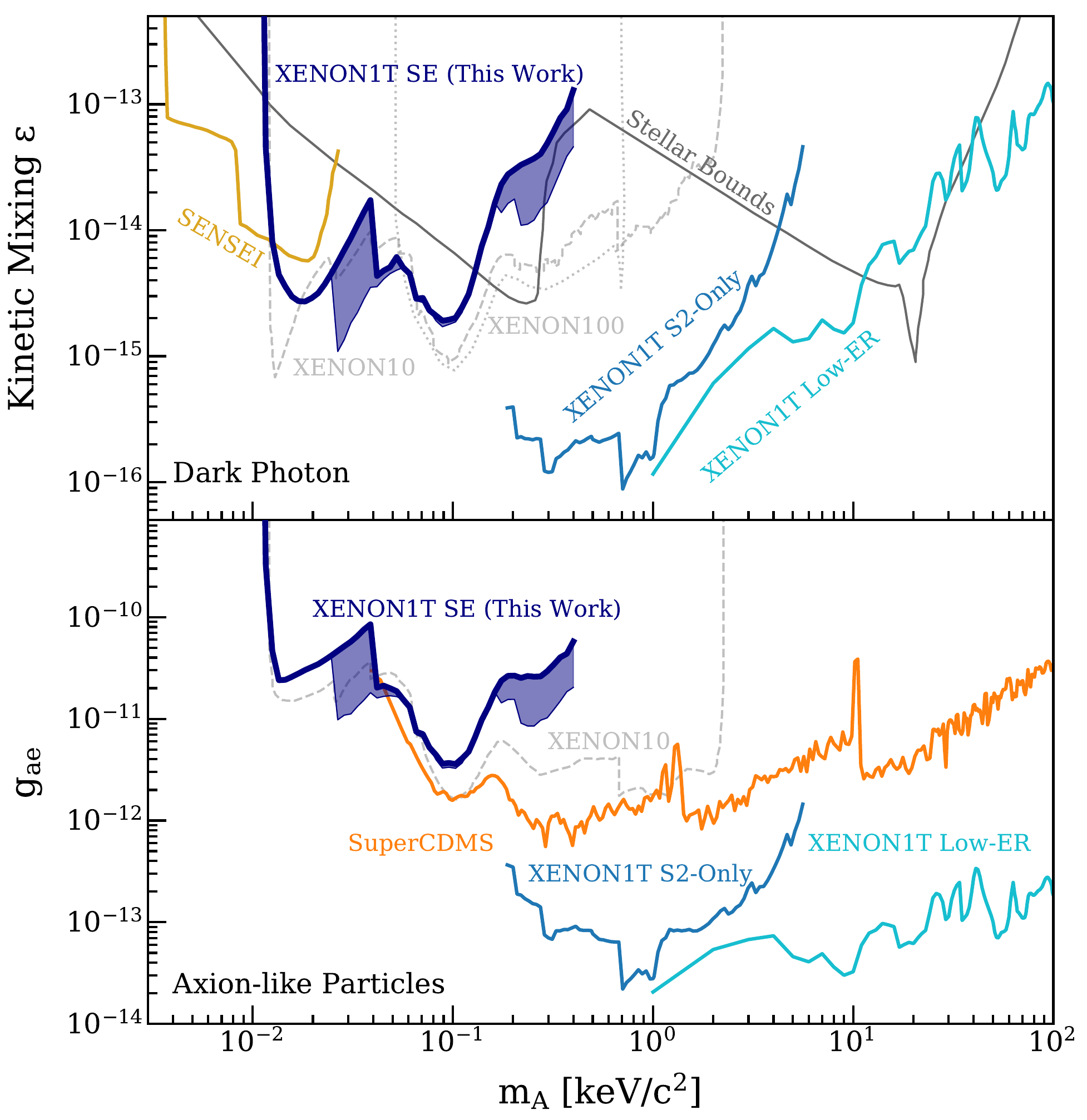}
    \caption{The 90\% confidence level upper limits on bosonic DM (dark blue) via dark photons (top) and ALPs (bottom), as function of DM mass $m_{A}$. The blue shaded band indicates the systematic uncertainty induced by the unknown differential ionization rate of the various electron shells in xenon. For comparison, we show experimental results (solid) from XENON1T S2-only~\cite{XENON:2019xxb} (light blue),
    XENON1T Low-ER~\cite{XENON:2020rca} (cyan),
    SENSEI~\cite{SENSEI:2020dpa} (gold), and
    SuperCDMS~\cite{SuperCDMS:2019jxx} (orange), alongside limits calculated (gray) in Ref.~\cite{Bloch:2016sjj} using data from XENON10~\cite{XENON10:2011prx} (dashed) and XENON100~\cite{XENON100:2013wdu} (dotted). Also shown are astrophysical constraints~\cite{An:2014twa} (dark gray).}
    \label{fig:dp_alp}
\end{figure}{}

\paragraph{\textbf{Bosonic Dark Matter}} Pseudo-scalar DM, such as axion-like particles (ALPs), or vector-boson DM candidates, such as dark photons, would be detectable through absorption by xenon atoms within the TPC. \textit{Dark photons} would be absorbed as a massive non-relativistic particle with monoenergetic signal at the mass of the dark photon, $m_\mathrm{A'}$, where the strength of the kinetic mixing between the photon and dark photon is given by $\epsilon$. \textit{Axion-like particles} (ALPs) interact with electrons through the ``axioelectric'' effect~\cite{Dimopoulos:1985tm}, where axions may be absorbed by bound electrons in the xenon atom, resulting in a monoenergetic signal at the rest mass, $m_\mathrm{A}$, of the particle. The absorption rate is dependent on the axion-electron coupling strength $g_\mathrm{ae}$. 

Exclusion limits are shown for dark photons and ALPs in the top and bottom panels of Figure~\ref{fig:dp_alp}, respectively. We report our limits assuming that ionized electrons are always produced from the lowest electron shell for which the mass of the DM particle exceeds the binding energy of that specific shell. This approach is more conservative than that adopted in Ref.~\cite{Bloch:2016sjj}, where the ionized electron is assumed to always originate from the outer most 5p electron shell. A complete analysis would require a careful treatment of the differential ionization rate for each shell. In order to compare directly to previous results, and to provide an estimate of the systematic uncertainty stemming from the unknown differential ionization rate, we also report our limit calculated under the less conservative assumption used in Ref.~\cite{Bloch:2016sjj}, where the uncertainty between the two assumptions is covered as a blue shaded region in Figure~\ref{fig:dp_alp}. Additionally limits from direct experimental results are shown in solid lines, and calculated limits in gray lines. In both cases, we probe lower mass ranges than previous XENON1T results, and exclude new parameter space for dark photons in a narrow mass range. 

\begin{figure}[t!]
    \centering
    \includegraphics[width=0.98\columnwidth]{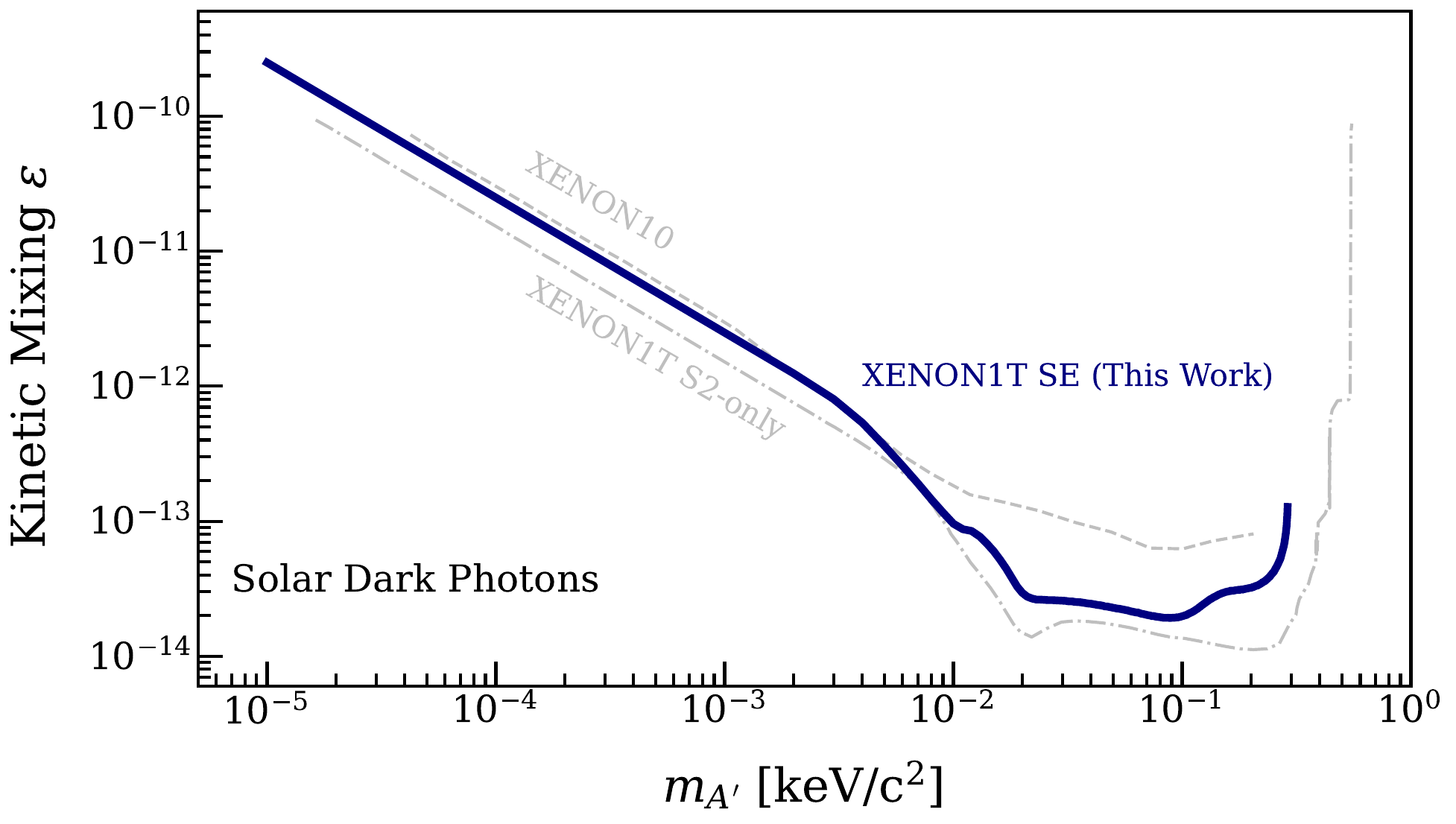}
    \caption{The 90\% confidence level upper limits on bosonic DM via solar dark photons (dark blue), as function of DM mass $m_\mathrm{A'}$. For comparison, we show limits calculated (gray) in Ref.~\cite{An:2013yua} using data from XENON10~\cite{XENON10:2011prx} (dashed) and Ref.~\cite{An:2020bxd} using data from XENON1T S2-only~\cite{XENON:2019xxb} (dot-dashed).}
    \label{fig:solar}
\end{figure}

\paragraph{\textbf{Solar Dark Photon}} Finally, we consider the case of dark photons originating in the Sun. The energy spectrum and flux of solar dark photons will differ greatly from relic DM dark photons as discussed in Appendix~\ref{sec:solar_dp}. The absorption rate of solar dark photons in LXe is strongly affected by their kinetic energy, which may be orders of magnitude higher than the rest energy, and the polarization, which is not isotropic, of the solar dark photon. The 90\% confidence upper limits for solar dark photons is presented in Figure~\ref{fig:solar}. Since the solar dark photons may be produced with considerable kinetic energy, the expected recoil spectrum is maximal in our 3-5 electron region, where we report a lower background rate in the 42-150\,PE (2-5 electrons) region relative to XENON10. Our work represents the first direct limits from experimental results on solar dark photons.

{}

%% file: sec_7_conclusion.tex
\section{Conclusion}
We have studied the background of single and few-electron S2 signals in XENON1T. This instrumental background has been observed in previous LXe TPCs and has presented an obstacle to push ionization-only searches for DM candidates in these detectors down to the lowest number of detected quanta. We attribute this background as originating from impurities within the LXe target volume. In doing so, we were able to develop data selection criteria to optimize the signal-to-background ratio in XENON1T for ionization signals which produce less than five electrons. 

Conservatively assuming that all remaining interactions observed are DM candidates, we set upper limits on the interactions for a number of DM models, excluding new parameter space for dark photons and solar dark photons. Despite our optimization, the instrumental background of 1 and 2 electron signals remains large in XENON1T. Future studies of delayed electron emission in upcoming next-generation detectors such as XENONnT~\cite{XENON:2020mcp}, LZ~\cite{LUX-ZEPLIN:2018poe, Akerib:2021pfd} and PandaX-4T~\cite{PandaX:2018wtu} are required to suppress the instrumental background for the smallest S2 signals (in terms of number of extracted electrons). 

With XENONnT having already achieved a LXe target $\mathcal{O}$(10) times more pure than XENON1T~\cite{PienaarJ} (as measured using the lifetime of drifting electrons), it is expected to be more sensitive to a number of the DM models presented here. The larger exposure  of XENONnT and its continuous data acquisition system can be leveraged to more accurately characterize this dominant background for sub-GeV DM searches in LXe TPCs.

%% file: sec_8_acknowledgements.tex
\section*{Acknowledgments}

We thank Kohsaku Tobioka and Tien-Tien Yu for useful discussions regarding dark photon interactions. We also thank Timon Emken for details regarding the calculation of signal spectra in effective theories. We gratefully acknowledge support from the National Science Foundation, Swiss National Science Foundation, German Ministry for Education and Research, Max Planck Gesellschaft, Deutsche Forschungsgemeinschaft, Helmholtz Association, Dutch Research Council (NWO), Weizmann Institute of Science, Israeli Science Foundation, Fundacao para a Ciencia e a Tecnologia, R\'egion des Pays de la Loire, Knut and Alice Wallenberg Foundation, Kavli Foundation, JSPS Kakenhi in Japan and Istituto Nazionale di Fisica Nucleare. This project has received funding/support from the European Union’s Horizon 2020 research and innovation programme under the Marie Sk\l odowska-Curie grant agreement No 860881-HIDDeN. Data processing is performed using infrastructures from the Open Science Grid, the European Grid Initiative and the Dutch national e-infrastructure with the support of SURF Cooperative. We are grateful to Laboratori Nazionali del Gran Sasso for hosting and supporting the XENON project.

%% file: appendixA.tex
\section{DM Models}

Candidate DM particles with sub-GeV masses can originate from dark-sector models such as those in Ref.~\cite{Arkani-Hamed:2008hhe, Cheung:2009qd}, that have a new $U(1)_D$ gauge group, independent from the Standard Model (SM). In these models, the interactions between DM and electrons are mediated by a dark-sector gauge boson coupling with charged SM particles via kinetic mixing with a photon~\cite{Holdom:1985ag}. A benchmark model is the minimal kinetically mixed dark photon Lagrangian~\cite{Alexander:2016aln}:

\begin{equation}\label{eq:minimal_l}
\mathcal{L}=-\frac{1}{4}F^{'\mu\nu}F'_{\mu\nu}+\frac{1}{2}\frac{\epsilon}{\cos\theta_\mathrm{W}}B^{\mu\nu}F^{'}_{\mu\nu}-\frac{1}{2}m_\mathrm{A'}^2A^{'\mu}A'_{\mu}
\end{equation}

\noindent with $A^{'}_{\mu}$ the dark photon vector field, $F^{'}_{\mu\nu}$ and $B_{\mu\nu}$ the field strength of the dark photon and of the SM hypercharge, respectively. The free parameters of this model are the dark photon mass, $m_\mathrm{A^{'}}$, and the kinetic mixing parameter, $\epsilon$. After electroweak symmetry breaking, a mixing with the SM electromagnetic field is introduced, resulting in a coupling $\epsilon e$ of the dark photon to electrically charged particles of the SM.

\subsection{Dark Matter-Electron Interactions}\label{sec:dm_e_scattering}
A fermionic, $\chi$, or a scalar boson, $\phi$, DM candidate can scatter off an electron bound to a xenon atom via a dark photon mediator. Following the model-independent approach of~\cite{Essig:2015cda}, we parameterize the elastic scattering cross section in the following way:

\begin{equation}\label{eq:cross_section}
\bar{\sigma}_\mathrm{e}=\frac{\mu_\mathrm{\chi,e}^2 }{16\pi m_\mathrm{\chi}^2 m^2_\mathrm{e}}|\mathcal{M}(q)|^2\Big|_{q=\alpha m_\mathrm{e}}
\end{equation}

\noindent with $\mathcal{M}(q)$ being the matrix element for elastic scattering of a free electron by a DM particle and $\mu_\mathrm{\chi,e}^2$ representing the reduced mass of the DM-electron system. Here, $\bar{\sigma}_\mathrm{e}$ is defined for a fixed momentum transfer of $q=\alpha m_\mathrm{e}$, with $\alpha$ denoting the fine-structure constant and $m_\mathrm{e}$ being the mass of the electron. The full momentum dependence is parameterized through the DM form factor, $F_{DM}(q)$ as $|\mathcal{M}(q)|^2=|\mathcal{M}(q)|^2|_{q=\alpha m_\mathrm{e}}\times |F_{DM}(q)|^2$. The precise functional form of $F_{DM}(q)$ depends on the specific interaction.

The velocity-averaged cross section for the ionization of an atomic electron, situated in the atomic orbital with quantum numbers $(n, l)$, can be written as:

\begin{equation}\label{eq:differential_cross_section}
\frac{d\langle\sigma_{\mathrm{ion}}^{nl} v\rangle}{d\ln E_\mathrm{r}}=\frac{\bar{\sigma_\mathrm{e}}}{8\mu_\mathrm{\chi,e}^2}\int q |f_{\mathrm{ion}}^{nl}(p',q) |^2 |F_{\mathrm{DM}}(q)|^2 \eta(v_{\textrm{min}}) dq
\end{equation}

\noindent where $\eta(v_{\mathrm{min}})$ is the mean-inverse speed and $v_{min}$ is the minimum speed required by a DM particle of mass, $m_{\chi}$, in order to ionize an electron from an initial bound state with binding energy $|E^{nl}_{\textrm{binding}}|$. 

In Eq.~\ref{eq:differential_cross_section}, the fact that the electron is in an atomic bound state is captured by the ionization form factor $|f_{\mathrm{ion}}^{nl}(p’,q)|$. For this function, we use the calculation provided in~\cite{yu}, which treats the initial and final states of the electron as bound and unbound states of the atomic system, respectively. The differential event rate is then proportional to the coherent sum of the thermally averaged differential cross section, for every atomic orbital:

\begin{equation}\frac{dR}{d\ln E_\mathrm{r}}=N_\mathrm{T}\frac{\rho_\mathrm{\chi}}{m_\mathrm{\chi}}\sum_{n,l}\frac{d\langle\sigma_{\mathrm{ion}}^{nl} v\rangle}{d\ln E_\mathrm{r}}
\end{equation}

\noindent where $\rho_\mathrm{\chi}$ is the local DM density and $N_\mathrm{T}$ is the number density of xenon atoms in the target. 

Ionization electrons with energy $E_\mathrm{r}$, can produce integer $n^{(1)}=(E_\mathrm{r}/W)$ additional quanta. Additionally, if the DM particle scatters off an electron of an inner shell, the photons produced by the resultant relaxation of the electron shell will produce integer $n^{(2)}=(E_i-E_j)/W$ secondary quanta, where $E_i-E_j=\Delta E$ is the difference between the binding energies of the initial and final states of the electron transition. We use the table given in~\cite{Essig:2017kqs} in order to estimate $n^{(2)}$. We express the probability of production of $n$ electrons due to the primary ionization electron, following \cite{Essig:2017kqs}, as: 

\begin{equation}
\begin{split}
P(n|n^{(1)},n^{(2)},p_\mathrm{qe},\langle r\rangle)=\langle r\rangle\text{Binom}(n|n^{(1)}+n^{(2)}, p_\mathrm{qe})\\+(1-\langle r\rangle)\text{Binom}(n-1|n^{(1)}+n^{(2)}, p_\mathrm{qe}) 
\end{split}
\end{equation}

\noindent where $p_\mathrm{qe}$ is the probability of observing excitation quanta as an electron.

\paragraph*{Effective Field Theory} In Eq.~(\ref{eq:cross_section}) the simplifying assumption is made that the scattering amplitude depends only on the transferred momentum. In Ref.~\cite{Catena:2019gfa} however, a non-relativistic effective theory is used in order to derive the most general form for this amplitude. The scattering amplitude can then be written as:

\begin{equation}\label{eq:eft_amplitude}
\mathcal{M}=\sum_{i}\Big(c_i^{s}+c_i^l\frac{q_{\textrm{ref}}^2}{|\mathbf{q}|^2}\Big)\langle\mathcal{O}_i\rangle
\end{equation}

\noindent with $q_{\textrm{ref}}=\alpha m_\mathrm{e}$ and $\langle\mathcal{O}_i\rangle$ the interaction operators given in Table I of Ref.~\cite{Catena:2019gfa}. In Eq.~\ref{eq:eft_amplitude}, $c_i^{s}$ and $c_i^l$ are the coupling constants corresponding to contact and long-range interactions respectively. In this framework, the aforementioned DM-electron scattering via heavy and ultra-light mediators are a special case of Eq.~\ref{eq:eft_amplitude} for the operator $\mathcal{O}_1$. 

\subsection{Bosonic Dark Matter}\label{sec:dm_bosonic}
It is possible for DM to consist entirely of pseudo-scalar or vector bosons, such as axion-like particles (ALPs) or dark photons, respectively. These particles can be detected through absorption by the atoms of the target medium, depositing their rest mass energy, and resulting in the production of ionization electrons~\cite{Bloch:2016sjj}. The absorption rate in a detector with a density of target atoms, $\rho_\mathrm{T}$, is given by~\cite{Hochberg:2016sqx}:

\begin{equation}\label{absorption_rate_general}
R=\frac{1}{\rho_\mathrm{T}}\frac{\rho_{\chi}}{m_{\chi}}\langle n_\mathrm{e}\sigma_{\mathrm{abs}}v_{\mathrm{rel}}\rangle
\end{equation}

\noindent where the factor $\langle n_\mathrm{e}\sigma_{\mathrm{abs}}v_{\mathrm{rel}}\rangle$ can be evaluated from the polarization tensor $\Pi$ using the optical theorem. The in-medium transverse $\Pi_\mathrm{T}$ and longitudinal $\Pi_\mathrm{L}$ modes of the polarization tensor are affected by the complex refractive index of the medium $n_{\mathrm{ref}}$~\cite{An:2014twa, Bloch:2016sjj}:

\begin{equation}
\Pi_\mathrm{L}=(\omega^2-\mathbf{q}^2)(1-n_{\mathrm{ref}}^2),\quad \Pi_\mathrm{T}=\omega^2(1-n_{\mathrm{ref}}^2)
\end{equation}

\noindent with $(\omega, \mathbf{q})$ being the 4-momentum of the absorbed particle. In the limit where the momentum, $\mathbf{q}$, is negligible, $\Pi_\mathrm{L}\approx\Pi_\mathrm{T} = \Pi(\omega)=\omega^2(1-n_{ref}^2)$, and consequently, $\langle n_\mathrm{e}\sigma_\mathrm{abs}v_\mathrm{rel}\rangle = -\textrm{Im}~ \Pi(\omega)/\omega$.

\paragraph{Dark Photon}
A dark photon $A'$ resulting from the broken gauge group $U(1)_D$ could constitute all relic DM as it would be stable on cosmological timescales if the kinetic mixing is sufficiently small and $m_{A'}<2m_\mathrm{e}$~\cite{Bloch:2016sjj}. The dark photon would be absorbed as a massive non-relativistic particle with monoenergetic signal at the rest mass, $m_{A'}$, with the absorption rate given by:

\begin{equation}
R=-\frac{1}{\rho_\mathrm{T}}\frac{\rho_\mathrm{\chi}}{m_\mathrm{A'}}\epsilon_{\textrm{eff}}^2\frac{\text{Im} ~\Pi(m_\mathrm{A'})}{m_\mathrm{A'}}
\end{equation}

\noindent where $\epsilon_{\textrm{eff}}$ is an effective mixing parameter. This is given by~\cite{Hochberg:2016sqx}:

\begin{equation}
\epsilon_{\textrm{eff}}^2=\frac{\epsilon^2 m_\mathrm{A'}^4}{(m_\mathrm{A'}^2-\textrm{Re}~\Pi(m_\mathrm{A'}))^2+(\textrm{Im}~\Pi(m_\mathrm{A'}))^2}
\end{equation}

\noindent and accounts for the modification of the kinetic mixing due to in-medium dispersion effects that would arise in LXe.

\paragraph{Axion-like Particles}
The axion, which was originally proposed as a solution to the strong CP problem~\cite{Peccei:1977hh}, is a pseudo-scalar particle that can couple to the SM axial current $\partial^{\mu}aJ^{A}_{\mu}/f_\mathrm{a}$, with constant $f_\mathrm{a}$ describing the scale at which the Peccei-Quinn symmetry is broken. In this model of axions, the mass $m_\mathrm{a}$ and the constant $f_\mathrm{a}$ are related by $m_\mathrm{a}=5.7~\textrm{meV}\times\frac{10^{9}~\textrm{GeV}}{f_\mathrm{a}}$. Other well-motivated models, for example ALPs which can be considered as low-energy remnants of discrete symmetries~\cite{Dias:2014osa}, can be described by independent parameters $m_\mathrm{a}$ and $f_\mathrm{a}$ while also coupling to gluons, photons and fermions. The minimal Lagrangian of this model can be written as:

\begin{equation}
\mathcal{L}=\frac{1}{2}\partial_{\mu}a\partial^{\mu}a-\frac{1}{2}m_\mathrm{a}^2a^2+ig_\mathrm{ae}a\bar{e}\gamma_{5}e
\end{equation}

\noindent where the dependence on $g_{ae}$ can be absorbed in $f_\mathrm{a}$ as $g_{ae}=2m_\mathrm{e}/f_\mathrm{a}$. ALPs are viable candidates for the DM relic density as they can be produced thermally or non-thermally, or through the misalignment mechanism~\cite{Arias:2012az}. ALPs interact with electrons through the ``axioelectric'' effect, where axions may be absorbed by bound electrons of the xenon atom resulting in a monoenergetic signal at the rest mass $m_\mathrm{a}$. For a xenon target, the absorption rate is given by:

\begin{equation}
R=-\frac{\rho_\mathrm{\chi}}{\rho_\mathrm{T}}\frac{3m_\mathrm{a}}{4m_\mathrm{e}^2}\frac{g_{ae}^2}{e^2}\frac{\textrm{Im}~\Pi(m_\mathrm{a})}{m_\mathrm{a}}
\end{equation}

\paragraph{Solar Dark Photon}\label{sec:solar_dp} 
We also consider the case of dark photons originating in the Sun rather than making up the bulk of the DM density. The energy spectrum and flux of solar dark photons will differ greatly from relic DM dark photons. To compute the flux rate for solar dark photons at a given mass and energy, we use theoretical models of the temperature and electron density in the solar interior. We also do not assume that the dark photon energy is dominated by its rest mass as dark photons from the Sun may have kinetic energy orders of magnitude higher than their rest mass. This will strongly affect the absorption rate of dark photons in LXe. Finally, unlike dark photons from relic DM, the polarization of dark photons from the Sun is not assumed to be isotropic as the production rates for longitudinally and transverse polarized dark photons differ. The absorption rates of these modes diverge from one another in cases where the energy of the dark photon is greater than its rest mass.

Considering these factors, we arrive at the equation for the event rate in LXe given by \cite{Bloch:2016sjj}:

\begin{equation}
R=\int_{m_{A'}}^{\omega_{\mathrm{max}}} \frac{\omega}{\sqrt{\omega^2-m_\mathrm{A'}^2}} \left[ \frac{d\Phi_\mathrm{T}}{d\omega}\Gamma_\mathrm{T}+\frac{d\Phi_\mathrm{L}}{d\omega}\Gamma_\mathrm{L} \right] d\omega
\end{equation}

\noindent where $\Phi_\mathrm{T}$ and $\Phi_\mathrm{L}$ represent the flux rates on Earth for transverse and longitudinal dark photons, respectively. The corresponding absorption rates, denoted by $\Gamma_\mathrm{T}$ and $\Gamma_\mathrm{L}$,  are dependent on the energy, $\omega$, as well as the dark photon mass, $m$, and coupling, $\epsilon$. We consider all possible energies of a dark photon with given mass $\omega=m_\mathrm{A'}$ up to the maximum plasma frequency in the Sun $\omega_{\mathrm{max}}\sim$293\,eV. Analogously to \cite{Bloch:2016sjj}, we determine the absorption rates to be: 

\begin{equation}
\Gamma_{\mathrm{T,L}}=-\epsilon_{\textrm{eff,T,L}}^2 \textrm{Im}~\Pi_{\mathrm{T,L}}{\omega}
\end{equation}

\noindent where $\epsilon_{\textrm{eff}}$ is modified such that $\Pi$ goes to $\Pi_\mathrm{L}(\omega)$ for longitudinally polarized dark photons and $\Pi_\mathrm{T}(\omega)$ for transverse polarized dark photons.

For the differential production rates of dark photons at a given energy in the Sun, we use the solar model from Ref.~\cite{Bahcall:2004pz}. We consider only the resonant production mode which dwarfs the other modes by about four orders of magnitude in our ROI. Longitudinally polarized dark photons will be resonantly produced in the Sun at a given radius, $r_\textrm{res}$, at which the plasma frequency, $\omega_\mathrm{p}$, is equal to the energy of the dark photon $\omega_\mathrm{p}=\omega$. Transverse polarized dark photons will be produced at the radius at which the plasma frequency equals the mass of the dark photon $\omega_\mathrm{p}=m_{A'}$.  The flux of solar dark photons is then given by Ref. \cite{An:2014twa} :

\begin{equation}
\frac{d\Phi_{\mathrm{T,L}}}{d\omega}=\frac{1}{4 \pi R_{\bigoplus}^2} \frac{2 \epsilon^2 r_{\mathrm{res}}^2}{e^{w/T({r_\textrm{res}})}-1}\frac{\sqrt{w^2-m_\mathrm{A'}^2}}{\sqrt{\frac{d\omega_\mathrm{p}^2}{dr}|_{r_\textrm{res}}}}
\end{equation}

\noindent where this rate is multiplied by a factor of $\omega^2 m_\mathrm{A'}^2$ in the longitudinal case and by $m_\mathrm{A'}^4$ for the transverse case. These production and absorption rates are combined and integrated across all relevant energies to get a total spectral event rate for dark photons of given mass in liquid xenon.

%% file: appendix_single_photons.tex
\section{Single Photoelectrons}

\begin{figure}[h]
    \centering
    \includegraphics[width=0.98\columnwidth]{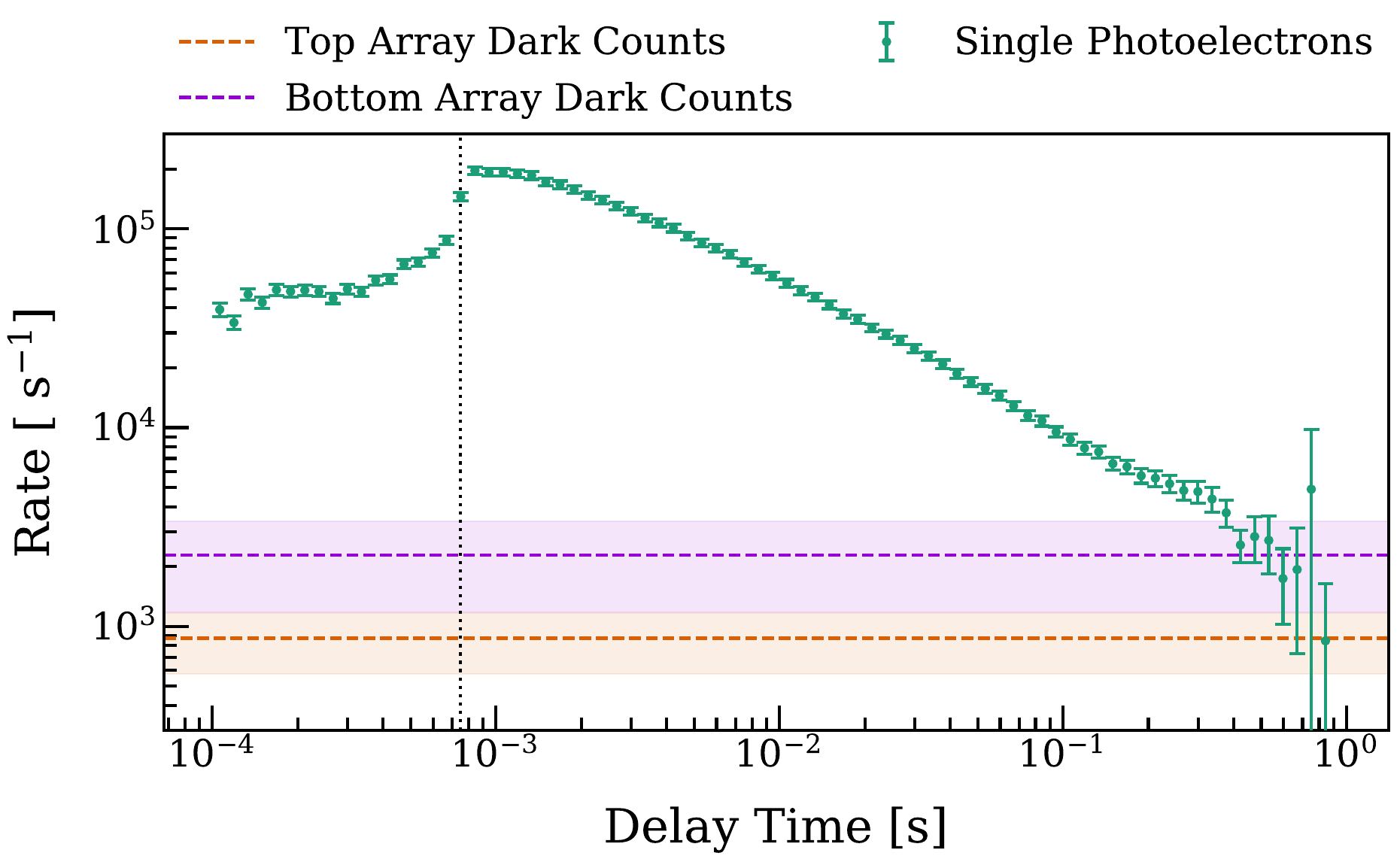}
    \caption{The rate of single photoelectron emission as a function of time following an energy deposition in the detector. A power-law dependence is observed, similarly to the delayed electron emissions. The dotted vertical line indicates the maximum drift time in XENON1T for SR1 and SR2. The dashed orange (purple) horizontal line indicates the median dark count rate in the top (bottom) PMT arrays used in XENON1T as was measured in Ref.~\cite{Rauchthesis}.}
    \label{fig:lonehit}
\end{figure} 

Previous studies~\cite{Akerib:2020jud, Sorensen:2017kpl} have observed an increased rate of single photoelectrons following a primary S2, and speculated that their origin could be photons from fluorescence in the PTFE material that lines the active region of the detector. While the focus of this work is understanding the delayed emission of few electron signals, we also investigate the time behavior of such single photoelectrons, isolated in time from any other interactions in the detector. As in the previous results, we observe an increased rate of single photoelectrons above the known dark count rate following a primary S2. Figure~\ref{fig:lonehit} shows the measured dark count rate~\cite{Rauchthesis} for the top (bottom) PMT array in XENON1T in the dashed orange (purple) line as well as the measured rate of single photons in the green data points. Similarly to the delayed electron emissions, the rate of single photoelectrons follows a decreasing power law. Here we find $\gamma\,=\,-0.7$, in contrast with the position correlated single electrons which have $\gamma\,=\,-1.1$. While these are single photoelectrons which are only detected by one PMT, the average amount of light observed in the top PMT array for these single photoelectrons, occurring 2-200\,ms after the primary S2, is consistent with the fraction of light observed in the top array for S1s produced in the active LXe between the cathode and the gate. This would disfavor PMT effects being the origin of this single photoelectrons background. Additionally, the spectrum of these single photoelectrons shows no evidence of double-photoelectron emission in the PMT, indicating that the origin is not from xenon scintillation light~\cite{Akerib:2020jud}.